\documentclass[12pt]{article}
\usepackage{latexsym}
\usepackage{amsmath,amsbsy,amssymb}

\usepackage{rotating}
\usepackage{curves}
\usepackage{epic}
\usepackage{eepic}
\usepackage{epsfig}

\newcommand{\dd}{\mbox{\rm d}}
\newcommand{\wg}{\wedge}
\newcommand{\gam}{\gamma}

\newcommand{\ddg}{\ddagger}
\newcommand{\tl}{\tilde}

\newcommand{\qb}{\qbezier}

\newcommand{\p}{\partial}

\newcommand{\be}{\begin{equation}}
\newcommand{\bear}{\begin{eqnarray}}

\newcommand{\ear}{\end{eqnarray}}
\newcommand{\ee}{\end{equation}}
\newcommand{\lbl}{\label}
\newcommand{\bi}{\bibitem}
\newcommand{\ci}{\cite}

\newcommand{\vs}{\vspace}
\newcommand{\hs}{\hspace}

\begin{document}

\begin{center}

\

\vs{.8cm}

\baselineskip .7cm

{\bf \Large Localized Propagating Tachyons 
in Extended Relativity Theories} \\

\vs{2mm}

\baselineskip .5cm
Matej Pav\v si\v c

Jo\v zef Stefan Institute, Jamova 39,
1000 Ljubljana, Slovenia

e-mail: matej.pavsic@ijs.si

\vs{3mm}

{\bf Abstract}
\end{center}

\baselineskip .43cm

\small We examine the possibility of localized  propagating tachyonic fields
within a properly extended relativity. A possible extension is to
include superluminal transformations and reference frames. This leads
to complex 4$D$ spacetime, or  real 8$D$ spacetime $M_{4,4}$.
The mass shell constraint in $M_{4,4}$ becomes, after first quantization,
the ultrahyperbolic Klein-Gordon equation. The Cauchy problem for such
equation is not well posed, because it is not possible to freely specify
initial data on a 7D hypersurface of $M_{4,4}$. We explicitly demonstrate
that it is possible to do it on a space-like 4-surface for bradyons,
and  on a time-like 4-surface for tachyons. But then the evolution of a
bradyonic field into the four time-like directions, or the ``evolution"
of a tachyonic field into the four space-like directions, is not uniquely
determined. We argue that this is perhaps no so bad, because in quantum field
theory (after second quantization) the classical trajectories of fields are not
determined anyway, and so it does not matter, if they are not completely
determined already in the first quantized theory. A next possible extension
of relativity is to consider 16$D$ Clifford space, $C$, a space whose
elements are oriented $r$-volumes, $r=0,1,2,3,4$ of real 4$D$ spacetime.
Then the evolution parameter can be associated  with an extra light-cone
coordinate, e.g., with the sum of the scalar and
the pseudoscalar coordinate, and initial data can be given on a light-like hypersurface,
in which case the Cauchy problem is well posed. This procedure
brings us to the Stueckelberg theory which contains localized propagating
tachyons in 4$D$ spacetime.

\vs{3mm}


\hs{7mm}

\baselineskip .55cm

\section{Introduction}

Recent experimental results\,\ci{Opera} have revived the interest in
superluminal particles, the so called tachyons. Later, an error
was found in that experiment, but this fact itself does not exclude the
possibility  of the existence of tachyons (to be eventually detected in
some other experiment), because the
theory of relativity can be suitably adapted or extended so to admit such
particles\,\ci{TachyonOld}--\ci{Antippa}. The possibility of superluminal
electromagnetic waves was considered in Ref.\,\ci{Rodrigues}.
But nowadays, it is commonly accepted that tachyonic
fields do not propagate with superluminal speeds\,\ci{Shay,Robinett},
because it is not possible to localize them on a given hypersurface
in the 4-dimensional spacetime, $M_{1,3}$. Thus, the
Cauchy data for such fields cannot be specified. Usually, by ``tachyonic
field" is understood a field that satisfies, e.g., the Klein-Gordon
equation with the opposite sign of mass square. But according to the
extended relativity\,\ci{PavsicExtend}--\ci{Antippa}
that takes into account
not only the subluminal, but also the superluminal Lorentz transformations,
the latter fields are not true tachyonic fields. Namely, under a superluminal
boost in the $x^1$-direction, the coordinates $x^2$, $x^3$ and
momenta $p^2$, $p^3$ become imaginary, and the Klein-Gordon equation
transforms into an equation that has the same form as the original
equation, in which $x^0$ and $x^1$ are interchanged. The hypersurface, on
which ``initial'' data can be specified, is now a time-like hypersurface
$x^1= constant$. If we keep the definition of velocity as the derivative of
the coordinate with respect to $x^0$, then the group velocity of the
field is given by the derivative of $p^0$ with respect to $p^1$, and it is
greater than the velocity of light. Such a field is the true tachyonic field.
It can be obtained from a bradyonic field by a superluminal transformation.
The fields that satisfy the Klein-Gordon equation in $M_{1,3}$, with the
opposite mass square, cannot be obtained from the bradyonic fields by a superluminal
transformation. Therefore, they are not the tachyonic fields of the extended
relativity, but are completely different objects that, according to
extended relativity, have nothing to do with tachyons. Only the fields that
can be obtained from the bradyonic fields by a superluminal transformations
can be treated as true tachyonic fields.

The imaginary unit occurring in superluminal transformations (SLT) imply
that spacetime is a complex space, $M_{1,3}\times \mathbb{C}$. 
We can avoid\footnote{In some literature (see for
example\,\ci{Time3D}--\ci{Time3D3}), imaginary coordinates, occurring
in superluminal transformations, are avoided by considering a 6-dimensional
spacetime, $M_{3,3}$, with 3 time-like and 3 space-like dimensions.
Then, under a SLT in the $x^1$-direction, the space-like coordinates
$x^2$, $x^3$ transform into the time-like coordinates $t^2$, $t^3$, and
vice versa.}
imaginary coordinates, if instead of the 4$D$ complex spacetime
we use the real 8$D$ spacetime $M_{4,4}$. The Klein-Gordon equation in such 8$D$
spacetime is ultrahyperbolic, and it describes bradyons and tachyons, depending
on the sign of the squared mass. Regardless of whether the field is bradyonic
or tachyonic, the Cauchy problem is not well posed, because no space-like
or time-like 7D hypersurface exist on which the initial data can be
arbitrarily specified. We show explicitly how the initial data can be
arbitrarily chosen on lower dimensional space-like or time-like
surfaces\footnote{Such a 4$D$ time-like surface in $M_{4,4}$, for a fixed extra
coordinate, corresponds to the time-like 3D hyperfurface, $x^1=constant$,
mentioned in the previous paragraph.}
 of dimensionality 4,
and point out that this could perhaps be sufficient for a consistent
physics of bradyonic and tachyonic fields in $M_{4,4}$. 

Another possibility is to consider a space of even higher dimensionality,
and take initial data on suitable light-like hypersurfaces in such space.
It is well-known\,\ci{CauchyLightLike} that the Cauchy problem for ultrahyperbolic
equations is well posed on light-like hypersurfaces. If the latter hypersurfaces
are orthogonal to our spacetime $M_{1,3}$, then we arrive at the generalized
Stueckelberg equation with a parameter $\tau$ that is invariant under
the Lorentz transformations in $M_{1,3}$. A Stueckelberg field
$\psi (\tau,x^\mu, \mbox{\footnotesize \it extra coordinates})$, $\mu=0,1,2,3$, contains
not only bradyons, but also tachyons, because it can be decomposed into
components with the 4-momenta $p_\mu$ that are not constrained,
and so both signs of $p^\mu p_\mu$ are allowed.

We consider a specific choice for a higher dimensional space, namely
the 16$D$ Clifford space, $C$, that has been extensively discussed in the
literature\,\ci{CliffSpace}--\ci{PavsicBook}. $C$ is a manifold whose tangent space
at any of its points is the Clifford algebra $Cl(1,3)$ of the spacetime
$M_{1,3}$. Mathematically, $C$ is the space of oriented $r$-volumes,
$r=0,1,2,3,4$. Physically, it can be interpreted as a configuration space,
associated with extended objects living in 4$D$ spacetime. In this paper
we point out, how the Clifford space $C$ leads to the Stueckelberg
theory, and thus to a consistent theory of tachyons. We also consider the
generalized Dirac equation in $C$, and show that it contains, as particular
cases, the tachyonic Dirac equation by Chodos et al.\,\ci{Chodos}, and
the bradyonic and tachyonic Dirac equations considered in
Ref.\,\ci{Jentschura2}. Many other
important features of Clifford space, such as description of spinors,
Kaluza-Klein theories, possible unification of interaction, etc.,
have already been described in Refs.\,\ci{CliffSpace}--\ci{PavsicBook}.

\section{The extended relativity and superluminal transformations
in $M_{1,3}$}

Special relativity can be extended\,\ci{PavsicExtend}--\ci{Antippa} so to incorporate
slower and faster than light particles, the so called {\it bradyons}, $B$,
and {\it tachyons}, $T$, and the {\it superluminal transformations}, SLT,
that transform $B$ into $T$, and vice versa. In a given reference frame,
$S$, a particle is observed as a bradyon, if its velocity is $v<c$, and
as a tachyon, if $v>c$. Under a superluminal transformation, the reference
frame $S$ transforms into $S'$. A particle that has $v<c$ in $S$, is observed
in $S'$ to have $v>c$, and vice versa. Whether a particle is bradyon or
tachyon depends on the reference frame from which it is observed.
Subluminal Lorentz transformations preserve the quadratic form
$\dd s^2 = \eta_{\mu \nu} \dd x^\mu \dd x^\nu$, so that $\dd s'^2 = \dd s^2$.
On the contrary, superluminal Lorentz transformations change the sign of
$\dd s^2$, so that $\dd s'^2 = -\dd s^2$. For the Minkowski metric tensor we
take $\eta_{\mu \nu}={\rm diag} (1,-1,-1,-1)$.

A superluminal transformation in the $x$-direction, of a position
4-vector $x^\mu \equiv (t,x,y,z)$, $\mu=0,1,2,3$,
is\,\ci{PavsicExtend}--\ci{Antippa}
\be
      t' = \frac{t+v x}{\sqrt{v^2-1}}~,~~~~~~x' = \frac{vt+x}{\sqrt{v^2-1}}~,~~~~~~
      y'=i y~,~~~~~~z'=iz
\lbl{2.1}
\ee
It satisfies $\dd t'^2 - \dd x'^2 - \dd y'^2 -\dd z'^2 = 
-(\dd t^2 - \dd x^2 - \dd y^2 -\dd z^2)$.
An imaginary unit $i$ occurs in the transformation of $y$ and $z$, otherwise
the quadratic form could not change its sign. While transformations
among subluminal reference frames can be formulated without reference
to complex numbers, this is not the case for transformations between
subluminal and superluminal reference frames.

It has been shown in Refs.\,\ci{PavsicExtend}--\ci{Recami1} that under a superluminal
Lorentz transformation, the velocity addition formula remains the same
as in the case of a subluminal Lorentz transformations.
From (\ref{2.1}) we have
\be
    \frac{\dd x'}{\dd t'} \equiv u' = \frac{v + u}{1 + v u}~, ~~~~
    u\equiv \frac{\dd x}{\dd t}
\lbl{2.1a}
\ee
which is the same equation that comes out from a subluminal Lorentz
transformation. Here $u$ is an object's velocity observed in $S$, and
$u'$ the velocity observed in $S'$, whereas $v$ is the relative velocity
between $S$ and $S'$. In Eq.\,(\ref{2.1a}) one can plug whatever velocities
$v \in (-\infty,\infty)$, $u \in (-\infty,\infty)$. As result one
obtains a corresponding velocity $u'$, that can be either subluminal or
superluminal. The case with $v \in (-1,1)$ and $u \in (1,3)$ has been
illustrated\footnote{
      Here the meaning of $u$ and $v$ is interchanged.}
in Fig.\,1 of Ref.\,\ci{Jentschura1}, where a 2D plot shows that the superluminal
velocity $u$, observed in $S$, remains superluminal upon transformation
with a subluminal velocity $v$ from a frame $S$ into a frame $S'$.   

The transformations, analogous to (\ref{2.1}), hold for the 4-momentum
$p^\mu = (p^0,p^1,p^2,p^3)  \equiv (p_t,p_x,p_y,p_z)$.
The component $p^0\equiv p_t \equiv E$ is a particle's energy.
Under a SLT we have
\be
     p'^\mu p'_\mu = -p^\mu p_\mu
\lbl{2.2}
\ee
If a particle is {\it bradyon} in $S'$, then it satisfies the mass shell
constraint
\be
    p'^\mu p'_\mu = m^2
\lbl{2.3}
\ee
Under a SLT, this becomes\
\be
    -p^\mu p_\mu = m^2
\lbl{2.4}
\ee
Here the proper mass, $m$, is assumed to be invariant under SLT.
More explicitly, a bradyon in $S'$, with the momentum
$(p'^0,p'^1,p'^2,p'^3)$, satisfies the mass shell constraint
\be
     (p'^0)^2 - (p'^1)^2 - (p'^2)^2 - (p'^3)^2 = m^2
\lbl{2.5}
\ee
Here $p'^0,p'^1,p'^2,p'^3$ are all {\it real}.
The same particle is observed in $S$ as a {\it tachyon} with momentum
$(p^0,p^1,p^2,p^3)$, satisfying the transformed mass shell
constraint
\be
     -(p^0)^2 + (p^1)^2 + (p^2)^2 + (p^3)^2 = m^2.
\lbl{2.6}
\ee

Assuming that $S$ and $S'$ are related by the superluminal transformation
(\ref{2.1}), we have that $p^0$, $p^1$ are real, while $p^2$, and $p^3$
are {\it imaginary}. Written in terms of real quantities,
$\tl p^2=i p^2$, $\tl p^3 = i p^3$,
 Eq.\,(\ref{2.6})
reads
\be
      (p^1)^2 - (p^0)^2 - (\tl p^2)^2 - (\tl p^3)^2 = m^2.
\lbl{2.7}
\ee
This is just like the usual, bradyonic, mass shell constraint, in
which $p^0$ and $p^1$ are interchanged. This fact has to be taken into
account when considering the Klein-Gordon equation for tachyons. Now $p^1$
has the same role as $p^0$ has in the bradyonic case. In other words,
the role of energy is now played by $p^1 \equiv p_x$:
\be
   p_x = \pm \sqrt{m^2 + p_t^2+{\tl p_y}^2+{\tl p_z}^2},
\lbl{2.7a}
\ee
where $\tl p_y \equiv \tl p^2,~\tl p_z \equiv \tl p^3$.

The Klein-Gordon equation in the reference frame $S'$ is obtained from
the constraint (\ref{2.5}), in which we replace momenta with
the operators $p'_\mu=-i \p'_\mu$, where $ \p'_\mu \equiv \p/\p x'^\mu$:
\be
     (-\p'_\mu \p'^\mu - m^2) \phi'(x') = 0.
\lbl{2.8}
\ee
A particular solution is
\be
    \phi'(x') = {\rm e}^{i p'_\mu x'^\mu} 
    = {\rm e}^{i(p'_t t'-p'_x x'-p'_y y'-p'_z z')  } .
\lbl{2.8a}
\ee
Under a SLT, we have
\be
   \phi'(x') = \phi(x)={\rm e}^{-i p_\mu x^\mu} 
    = {\rm e}^{-i(p_t t-p_x x-p_y y-p_z z)  } =
    {\rm e}^{-i(p_t t-p_x x+\tl p_y \tl y+\tl p_z \tl z)  }
\lbl{2.9}
\ee
In the last step we have taken into account that $p_y,~p_z,~y,~z$ are
imaginary, and expressed them in terms of the real quantities
$\tl p_y=i p_y,~\tl p_z=i p_z,~\tl y= iy ,~\tl z=i z$.

In the reference frame $S$ we have the Klein-Gordon equation, obtained from
the constraint (\ref{2.6}), which is equivalent to the constraint (\ref{2.7}):
\bear
    (\p_\mu \p^\mu - m^2) \phi(x) &=&
    \left ( \frac{\p^2}{\p t^2}- \frac{\p^2}{\p x^2}-\frac{\p^2}{\p y^2}
     -\frac{\p^2}{\p z^2} - m^2 \right ) \phi (t,x,y,z) \nonumber \\
   &=& \left ( \frac{\p^2}{\p t^2}- \frac{\p^2}{\p x^2}+\frac{\p^2}{\p \tl y^2}
     +\frac{\p^2}{\p \tl z^2} - m^2 \right ) \phi (t,x, \tl y,\tl z) = 0.
\lbl{2.10}
\ear
From the mathematical point of view, this is just like the usual
Klein-Gordon equation, with $t$ and $x$ interchanged.
A particular solution of the latter equation is (\ref{2.9}).

A general solution is
\be
    \phi = \int \dd p_t \, \dd p_x \,\dd \tl p_y \,\dd \tl p_z \,
    c(p_t,p_x,\tl p_y,\tl p_z) {\rm e}^{i (p_x x-p_t t-\tl p_y \tl y-\tl p_z \tl z)}
    \delta (p_x^2-p_t^2-{\tl p_y}^2-{\tl p_z}^2 - m^2) ,
\lbl{2.11}
\ee
where $c(p_t,p_x,\tl p_y,\tl p_z)$ is a function of its arguments, restricted
to the mass shell.
Introducing $\omega_x=\Bigl\vert \sqrt{m^2+p_t^2+{\tl p_y}^2+{\tl p_z}^2}\Bigl\vert$,
and integrating over $p_x=\pm \omega_x$, we obtain
\be
   \phi = \int \dd p_t \,\dd \tl p_y \,\dd \tl p_z \, \frac{1}{2 \pi \omega_x}
   \left [ {\rm e}^{i (\omega_x x - p_t t-\tl p_y \tl y -\tl p_z \tl z)}
   c(\omega_x,p_t,\tl p_y,\tl p_z) \right .  \hs{2cm}
\nonumber
\ee
\be
    \hs{3.5cm} +\left . {\rm e}^{i (-\omega_x x - p_t t-\tl p_y \tl y -\tl p_z \tl z)}
   c(-\omega_x,p_t,\tl p_y,\tl p_z) \right ] 
\lbl{2.12}
\ee
The ``initial" data can be given on a surface $x=x_0={\rm constant}$. The latter
surface is time-like, and is spanned by coordinates $(t,\tl y,\tl z)$.
Specifying ``initial" data on a hypersurface $x=x_0$, one can calculate
the field $\phi$ at other hypersurfaces, with different values of $x$.
That a Cauchy surface for space-like states must be time-like, was discussed
in Ref.\,\ci{Shay,BarutTachyon}, but without employing superluminal
transformations, and in Ref.\,\ci{Vysin,Vysin1}, within the framework
of 2-dimensional spacetime.

Let us now investigate how does move the wave packet that satisfied the
Klein-Gordon equation. For this purpose it is convenient to consider
the nonrelativistic approximation to the Klein-Gordon equation.
Writing a solution of Eq.\,(\ref{2.8}) as
$\phi'={\rm e}^{-i m t'} \psi' (t',{\bf x'})$, and substituting it into
Eq.\,({\ref{2.8}), we obtain\,\ci{Klein-Gordon}
\be
   -\frac{1}{2m} \, \frac{\p^2 \psi'}{\p t'^2} + i \frac{\p \psi'}{\p t'}
   =-\frac{1}{2m} \, \nabla'^2 \psi' .
\lbl{R1}
\ee   
In the nonrelativistic limit we can neglect the first term, and we obtain
the Schr\"odinger equation
\be
    i \frac{\p \psi'}{\p t'} = -\frac{1}{2m} \, \nabla'^2 \psi' .
\lbl{R2}
\ee
A solution to the latter equation is a Gaussian wave packet, whose square
is
\be
   |\psi'|^2 \propto {\rm exp} \left [-\frac{({\bf x'}
     -\frac{{\bf p'}_0}{m} t')^2}{\sigma' (t')} \right ]
\lbl{R3}
\ee
where $\sigma' (t') = \sigma'_0 + t'^2/(m^2 \sigma'_0)$. The probability
density is maximal on the classical trajectory ${\bf x'}=({\bf p'}_0/m) t'$
with velocity ${\bf v'}={\bf p'}_0/m$, $|{\bf v'}| < c=1$.
We have verified that the Klein-Gordon
Eq.\,(\ref{2.8}) in the reference frame $S'$ describes a field, propagating
with a slower than light velocity ${\bf v'}$ (Fig.\,1a).

The same procedure can be repeated for the Klein-Gordon equation (\ref{2.10})
that holds in the reference frame $S$, and in which $t$ and $x$ are
interchanged. Writing $\phi = {\rm e}^{-i m x} \psi$, and substituting this
into Eq.\,(\ref{2.10}), we obtain
\be
   - \frac{1}{2m} \, \frac{\p^2 \psi}{\p x^2} + i \, \frac{\p \psi}{\p x}
   = - \frac{1}{2m} \, \left ( \frac{\p^2}{\p t^2} + \frac{\p^2}{\p \tl y^2} 
   +\frac{\p^2}{\p \tl z^2} \right ) \psi
\lbl{R4}
\ee
In the approximation, in which the first term can be neglected, we have
\be
    i \, \frac{\p \psi}{\p x}
   = - \frac{1}{2m} \, \left ( \frac{\p^2}{\p t^2} + \frac{\p^2}{\p \tl y^2} 
   +\frac{\p^2}{\p \tl z^2} \right ) \psi
\lbl{R5}
\ee
Such equation, but with the $x$ and $t$ variables only, was considered
by Vy\v s\'in\,\ci{Vysin,Vysin1}.
A solution of the Schr\"odinger-like equation (\ref{R5}) is the wave
packet whose square is
\be
   |\psi|^2 \propto {\rm exp} \left [-\frac{(t-\frac{p_{t 0}}{m}x)^2}{\sigma (x)}
   -\frac{(\tl y-\frac{\tl p_{y 0}}{m}x)^2}{\sigma (x)} 
   -\frac{(\tl z-\frac{p_{{\tl z}0}}{m}x)^2}{\sigma (x)}
      \right ] ,
\lbl{R6}
\ee
where $\sigma(x)=\sigma_0 + x^2/(m^2 \sigma_0)$. The probability density
is now picked on the classical trajectory, $t=\frac{p_{t 0}}{m}x$, 
$\tl y-\frac{\tl p_{y 0}}{m}$, $\tl z-\frac{\tl p_{z 0}}{m}$,
where $p_{t 0}/m=\dd t/\dd x$, $\tl p_{y 0}/m=\dd \tl y/\dd x$, and
$\tl p_{z 0}/m=\dd \tl z/\dd x$, $|\sqrt{(\dd t/\dd x)^2 + (\dd \tl y/\dd x)^2 
+(\dd \tl z/\dd x)^2 }| < c=1.$
If we keep on considering $t$ as a time coordinate, then the velocity of
the wave packet (\ref{R6}) in the $x$-direction should be defined according
to $v_x=\dd x/\dd t = m/p_{t 0}$, where $|v_x| >1$.
The reciprocity between time and space for bradyons and tachyons was
noticed long time ago\,\ci{Antippa,Mignani}, and localization of tachyons
on the basis of such reciprocity was considered by Vi\v syn\,\ci{Vysin,Vysin1},
\ci{Shay}, and others\,\ci{TachyonLocal}--\ci{TachyonLocal2}.

\setlength{\unitlength}{.8mm}

\begin{figure}[h!]
\hs{3mm} \begin{picture}(120,70)(-7,-20)

\newsavebox{\bradyon}
\savebox{\bradyon}(40, 32)[l]{
\put(0,0){\epsfig{file=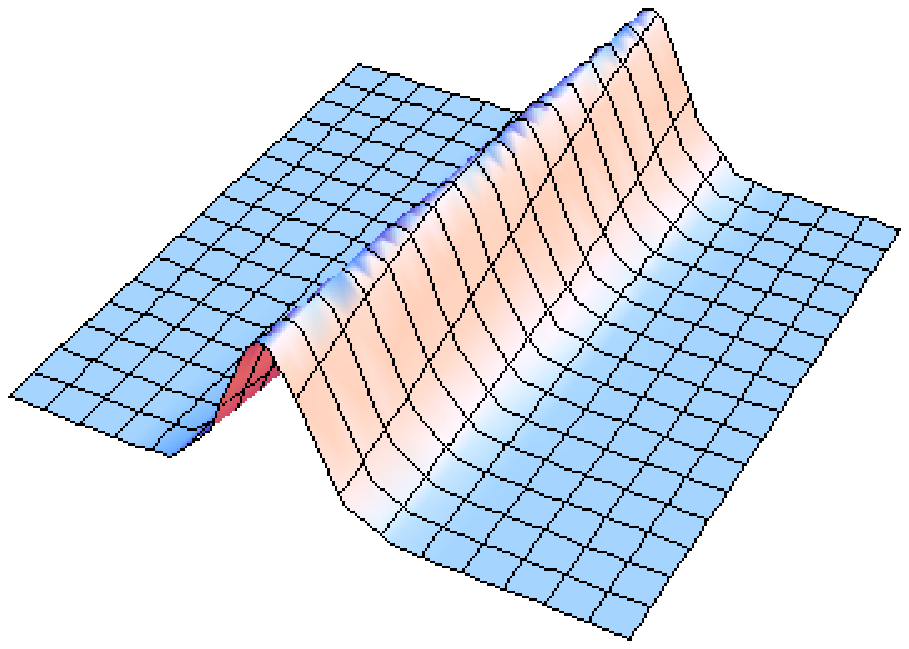,width=60mm}}
\put(57,-0.7){$x$}
\put(0,44.5){$|\psi|^2$}
\put(34.5,49){$t$}

\curve(4.5,19.5, 1.5,40.5)
\put(1.5,40.5){\vector(-1,4){0.2}}

\curve(4.5,19.5, 55,-0.7)
\put(55,-0.7){\vector(2,-1){1}}
\curve(4.5,19.5, 33.5,47)
\put(33.5,47){\vector(1,1){1}}

\put(33,-7){a)}
}

\newsavebox{\tachyon}
\savebox{\tachyon}(40, 32)[l]{
\put(0,0){\epsfig{file=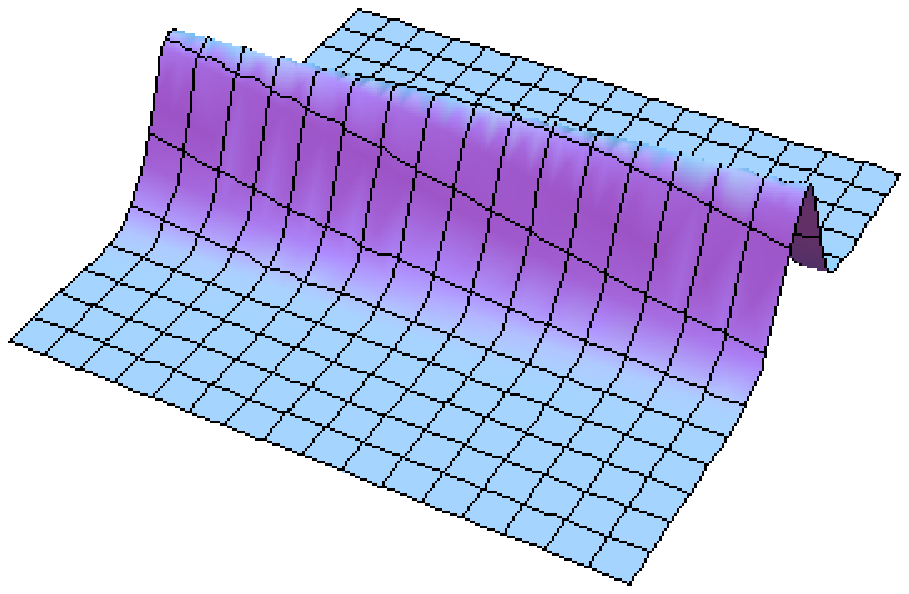,width=60mm}}
\put(57,-0.7){$x$}
\put(0,44.5){$|\psi|^2$}
\put(34.5,49){$t$}

\curve(4.5,19.5, 1.5,40.5)
\put(1.5,40.5){\vector(-1,4){0.2}}
\curve(4.5,19.5, 55,-0.7)
\put(55,-0.7){\vector(2,-1){1}}
\curve(26,39.88, 33.5,47)
\curve(4.5,19.5, 12,26.61)

\put(33.5,47){\vector(1,1){1}}

\put(33,-7){b)}
}

\put(0, 0){\usebox{\bradyon}}
\put(90,0){\usebox{\tachyon}}

\end{picture}

\caption{\footnotesize Examples of bradyonic (a) and tachyonic (b)
wave packets.}
\end{figure} 

We see that the wave packet (\ref{R6})
propagates with the superluminal velocity (Fig.\,1b). Because the subluminal
wave packet (\ref{R3}) can transmit a signal, so can the wave packet
(\ref{R6}), which is nothing but a superluminal transform of (\ref{R3}).

We illustrated the situation by considering the nonrelativistic wave packet.
In the relativistic case, we can use the dispersion relation (\ref{2.7a}).
We find that
\be
    \frac{\dd p_x}{\dd p_t} = \frac{\dd t}{\dd x}
    =\frac{p_t}{\sqrt{p_t^2 + \tl p_y^2 + \tl p_z^2 + m^2}}~,~~~~~~
    \Bigl\vert \frac{\dd t}{\dd x} \Bigl\vert < c=1
\lbl{R7}
\ee
gives the reciprocal group velocity of the tachyonic field in the
$x$-direction. It follows
that $\dd x/\dd t >1$. For simplicity, we now omit the subscript `$0$',
denoting the center of the wave packet.
That a tachyonic field propagates with a superluminal speed directly follows
from the relativistic dispersion relation $p_t= \sqrt{p_x^2-\tl p_y^2 -
\tl p_z^2 -m^2}$ that comes from Eq.\,(\ref{2.7}), and from which we obtain
the faster-than-light group velocity 
\be
    \frac{\dd p_t}{\dd p_x} = \frac{\dd x}{\dd t}
    =\frac{p_x}{\sqrt{p_x^2 - \tl p_y^2 - \tl p_z^2 - m^2}}~,~~~~~~ 
    \Bigl\vert \frac{\dd x}{\dd t} \Bigl\vert> c=1
\lbl{R8}
\ee

The {\it nonrelativistic} (i.e., very fast moving) tachyon wave packet is
localized in time, and also in space. From Fig.\,1b and Eq.\,(\ref{R6})
we see that the localization width in space is wide in comparison with
that in time. This is a non relativistic approximation. A {\it relativistic}
tachyon can be localized only in time,
but not in space. The fact that tachyons are infinitely extended in space,
was taken as an argument against
the possibility that they can transmit information. However, as
already observed by Vy\v s\'in\,\ci{Vysin,Vysin1}, it is important that
tachyons form sharp pulses in time. A sequence of such pulses, localized in time,
can encode information. If there is an interaction between tachyons and
a bradyonic detector, then a bradyonic observer $B$ would be able to
observe that sequence of pulses, and interpret it as the information,
emitted by another observer $A$. This is illustrated in Fig.\,2.

\setlength{\unitlength}{.8mm}

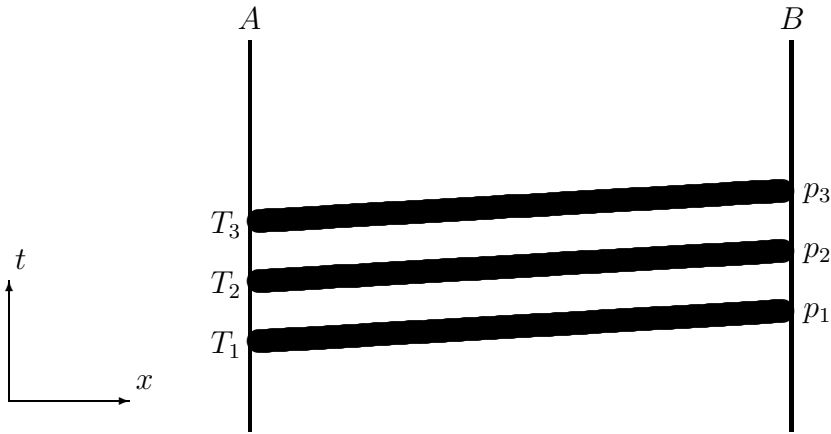
\begin{figure}[ht]
\hs{3mm} \begin{picture}(130,77)(-100,-20)

\put(-90,-10){\vector(1,0){20}}
\put(-90,-10){\vector(0,1){20}}
\put(-69,-8){$x$}
\put(-89,12){$t$}

\put(-56.5,-2){$T_1$}
\put(-56.5,8){$T_2$}
\put(-56.5,18){$T_3$}

\put(42,4){$p_1$}
\put(42,14){$p_2$}
\put(42,24){$p_3$}

\put(-52,52){$A$}
\put(38,52){$B$}

\linethickness{3mm}
\curve(-48.5,0, 38.5,5)
\curve(-48.5,10, 38.5,15)
\curve(-48.5,20, 38.5,25)

\linethickness{0.5mm}
\put(-50,-15){\line(0,1){65}}
\put(40,-15){\line(0,1){65}}

\end{picture}

\caption{\footnotesize  
Though the tachyons wave packets are not localized in the $x$-direction,
having no sharp peaks and leading edges along the $x$-direction, they can
nevertheless transmit information
from $A$ to be $B$, because they are localized in time. Namely,
if tachyons can interact with bradyons,
then the observer $A$ can send a message encoded in a sequence of
emitted tachyons, $T_1,~T_2,~T_3, ...$, that are
detected by the observer $B$ as a sequence of pulses, $p_1,~p_2,~p_3,...$, at a fixed
spatial position. $B$ can unambiguously interpret such sequence as a message,
if both observers had already agreed about the code.
}

\end{figure}

The solution (\ref{2.11}) is the superluminal transform of the general
solution of the Klein-Gordon equation (\ref{2.8}) for a bradyonic field
in $S'$.
Under a SLT, the bradyonic Klein-Gordon equation transforms into the
tachyonic Klein-Gordon equation. But according to the principle of relativity,
the laws of motion, encrypted in the equations of motion, should remain
unchanged under all transformations that bring one dynamically possible
solution into another dynamically possible solution. According to our
assumption, SLT are such transformations. Therefore, the equations of
motion should remain invariant under SLT. Since Eq.\,(\ref{2.8}) is not
invariant, it means that it is not a complete equation, but a part of a
more general equation. In the following we will consider a more general theory
which  the Klein-Gordon equation\,(\ref{2.10}) is embedded in. First,
we will discuss a generalization of the classical theory of the relativistic
point particle, and then its quantization.

\section{Complex coordinates and momenta, and the real spacetime $M_{4,4}$}

A superluminal boost (\ref{2.1}) contains imaginary quantities.
After performing, successively, a superluminal boost, rotations and
a subluminal boost, we will end up with complex spacetime coordinates
and momenta:
\be
   X^\mu = x^\mu + i \tl x^\mu~,~~~~~~P^\mu = p^\mu +i \tl p^\mu ,
\lbl{2.13}
\ee
with
\be
   X^\mu X_\mu = X^\mu X^\nu \eta_{\mu \nu} =
    x^\mu x_\mu - \tl x^\mu \tl x_\mu + 2 i x^\mu \tl x_\mu ,
\lbl{2.14}
\ee
\be
   P^\mu P_\mu = P^\mu P^\nu \eta_{\mu \nu} =
    p^\mu p_\mu - \tl p^\mu \tl p_\mu + 2 i p^\mu \tl p_\mu ,
\lbl{2.14a}
\ee
Subluminal Lorentz transformations preserve $X^\mu X_\mu$, so they
preserve $x^\mu x_\mu - \tl x^\mu \tl x_\mu$ and $x^\mu \tl x_\mu$.
Superluminal Lorentz transformations change the sign of $X^\mu X_\mu$.
A superluminal boost in the $x$-direction reads
\be
    T'= \frac{T+v X}{\sqrt{v^2-1}}~,~~~~X'=\frac{vT+X}{\sqrt{v^2-1}}~,~~~~
     Y'=iY~,~~~~Z'=i Z ,
\lbl{2.15}
\ee
where $v$ is real. For the real and imaginary components, we have
\be
    t'= \frac{t+v x}{\sqrt{v^2-1}}~,~~~~x'=\frac{vt+x}{\sqrt{v^2-1}}~,~~~~
     y'=-\tl y~,~~~~z'=-\tl z ,
\lbl{2.16}
\ee
\be
    \tl t'= \frac{\tl t+v \tl x}{v^2-1}~,~~~~\tl x'=\frac{v \tl t+\tl x}{v^2-1}~,
      ~~~~  \tl y'=y~,~~~~\tl z'=z 
\lbl{2.17}
\ee

Analogous transformations hold for the momentum $P^\mu$. In this setup with
complex coordinates and momenta, the multiplication by $i$ in the SLT
(\ref{2.15}) means the interchange of the real and imaginary components.

If in the reference frame $S'$, the imaginary components are zero,
$\tl x'^{\,\mu}=0$, $\tl p'^{\,\mu}=0$, then in $S$, according to Eqs.\,(\ref{2.16}),
(\ref{2.17}), we have
\be
   y=0~,~~~z=0~,~~~\tl y=-y'~,~~~\tl z=-z'
\lbl{2.18}
\ee
\be
   p^2=0~,~~~p^3=0~,~~~\tl p^2=-p'^{\,2}~,~~~\tl p^3=-p'^{\,3}
\lbl{2.19}
\ee
The quadratic forms read
\bear
   X'^{\,\mu} X'_\mu&=& x'^{\,\mu} x'_\mu=t'^{\,2}-x'^{\,2}-y'^{\,2}-z'^{\,2}
   \nonumber\\
    &=& -t^2+x^2-y'^{\,2}-z'^{\,2}= -t^2+x^2-\tl y^2-\tl z^2
\lbl{2.20}
\ear
\bear
   P'^{\,\mu} P'_\mu &=&p'^{\,\mu} p'_\mu 
   =({p'^{\,0}})^2-({p'^{\,1}})^2-({p'^{\,2}})^2-({p'^{\,3}})^2
   \nonumber\\
    &=& -({p^0})^2+({p^1})^2-({p'^{\,2}})^2-({p'^{\,3}})^2
    = -({p^0})^2+({p^1})^2-({\tl p^2})^2-({\tl p^3})^2 \hs{1cm}
\lbl{2.21}
\ear
This is precisely the quadratic form of Eq.\,(\ref{2.7}) that we arrived at
in Sec.\,2. The derivation through Eqs.\,(\ref{2.1})--(\ref{2.12}) belongs to
a special case of a more general theory in which coordinates and momenta
are complex.

Let  the momentum constraint in the reference frame $S'$ be
\be
   {\rm Re}\, P'^\mu P'_\mu +M^2 = 0
\lbl{2.22}
\ee
Assuming real $M^2$, we thus have
\be
   p'^\mu p'_\mu - \tl p'^\mu \tl p'_\mu +M^2 = 0
\lbl{2.23}
\ee
In a reference frame $S$, related to $S'$ by the SLT (\ref{2.16}),(\ref{2.17}),
we have
\be
   -({p^0})^2+({p^1})^2+({p^2})^2+({p^3})^2 
   +({\tl p^0})^2-({\tl p^1})^2-({\tl p^2})^2-({\tl p^3})^2+M^2 = 0 ,
\lbl{2.25}
\ee
i.e., 
\be
   -p^\mu p_\mu + \tl p^\mu \tl p_\mu +M^2 = 0
\lbl{2.26}
\ee
Here, again $M^2$ is assumed to be invariant under SLT. It may be positive,
or negative:

a) If $M^2 > 0$, then the particle is observed as a tachyon in $S'$,
and as a bradyon in $S$.

b) If $M^2 <0$, then the particle is observed as a bradyon in $S'$, and
as a tachyon in $S$.

Instead of the 4-dimensional complex spacetime $M_{1,3}\times \mathbb{C}$,
we can consider the 8-dimensional real space $M_{4,4}$ with signature $(4,4)$.

Upon quantization, the constraint (\ref{2.26}) becomes the Klein-Gordon
equation:
\be
    (\p^\mu \p_\mu - \tl \p^\mu \tl \p_\mu +M^2) \phi(x^\mu,\tl x^\mu)=0 ,
\lbl{2.26a}
\ee
where $\tl \p^\mu \tl \p_\mu=\tl \p^\mu \tl \p^\nu \eta_{\mu \nu}$, and
$\tl \p_\mu \equiv \p/\p \tl x^\mu$.

A general solution of this equation is
\be
   \phi=\int \dd^4 p \, \dd^4 \tl p \,c(p,\tl p) 
   {\rm e}^{i(p_\mu x^\mu - \tl p_\mu \tl x^\mu)}
    \delta (p^\mu p_\mu - \tl p^\mu \tl p_\mu -M^2)
\lbl{2.27}
\ee
Instead of one, we have now four time-like and four space-like coordinates
and momenta:
\bear
&&(x^0,\tl x^1,\tl x^2,\tl x^3);~~~(p^0,\tl p^1,\tl p^2,\tl p^3)
~~~~~\mbox{time-like}\nonumber \\
&&(x^1,x^2,x^3,\tl x^0);~~~(p^1,p^2,p^3,\tl p^0)
~~~~~\mbox{space-like}\nonumber
\ear

In the case of {\it the bradyonic field}, $M^2>0$, we can proceed as
follows. We choose the time-like component $p^0$ and integrate it out in
Eq.\,(\ref{2.27}). Denoting ${\bf x}\equiv (x^1,x^2,x^3)$, etc., we obtain:
   $$\phi(x^\mu,\tl x^\mu)= \int \dd^3 {\bf p}\, \dd \tl p^0\,\dd^3 {\bf \tl p}\,
   \frac{1}{2 \omega}\, \left [ {\rm e}^{i (\omega x^0- {\bf p}{\bf x}
   -\tl p_0 \tl x^0 +{\bf \tl p} {\bf \tl x})} 
   c(\omega, {\bf p}, \tl p^0,{\bf \tl p})
   \right .$$
\be
    \hs{3cm} \left . + \; {\rm e}^{i (-\omega x^0- {\bf p}{\bf x}
   -\tl p_0 \tl x^0 +{\bf \tl p} {\bf \tl x})} 
   c(-\omega, {\bf p}, \tl p^0,{\bf \tl p}) \right ] ,
\lbl{2.28}
\ee
where
\be
    p^0=\pm \omega~,~~~~~\omega =\vert \sqrt{(\tl p^0)^2+{\bf p}^2 -{\bf \tl p}^2
    +M^2} \vert
\lbl{2.29}
\ee
At the (7-dimensional)
hypersurface $x^0=0$, the initial data $\phi \vert_{x^0=0}$ and
$(\dd \phi/\dd x^0) \vert_{x^0=0}$ cannot be arbitrarily chosen.
Because the momenta, assumed to be real, are constrained by Eq.\,(\ref{2.29}), the expression
(\ref{2.28}) does not give arbitrary $\phi \vert_{x^0=0}$. Similarly for
the $x^0$-derivative of $\phi$. The Cauchy problem is thus not
well posed\,\ci{Courant}. 

This is the notorious property of ultrahyperbolic partial differential equations
that involve more than one time-like dimension. Usually, this is considered
as a problem, and it is concluded that ultrahyperbolic equations, such as
(\ref{2.26a}), are not suitable for physics. I think that the requirement
that initial data should be given arbitrarily on a hypersurface (which in
our case is a 7-surface), is an unnecessary restriction
on which equations are admissible in physics. We will discuss this in more
detail later. In the case of Eq.\,(\ref{2.26a}), the initial data can be
freely specified on the 4-surface, spanned by coordinates 
$(x^1,x^2,x^3,\tl x^0)$. This follows immediately from Eq.\,(\ref{2.28}).
Putting $x^0=0$, ${\bf \tl x}={\bf 0}$, we have
   $$\phi(0,{\bf x},\tl x^0,{\bf 0})= \int \dd^3 {\bf p}\, \dd \tl p^0\,\dd^3
    {\bf \tl p}\,
   \frac{1}{2 \omega}\, {\rm e}^{i (- {\bf p} {\bf x}-\tl p_0 \tl x^0 )} 
   \left [ c(\omega,{\bf p},\tl p^0, {\bf \tl p})
   +c(-\omega,{\bf p},\tl p^0, {\bf \tl p}) \right ] $$
\be
    =\int_{-\infty}^{\infty} \dd^3 {\bf p}\, \dd \tl p^0 \,  A({\bf p},\tl p^0)
     {\rm e}^{-i ({\bf p} {\bf x} +\tl p_0 \tl x^0 )} , \hs{3.2cm} 
\lbl{2.29a}
\ee
where
\be
    A({\bf p},\tl p_0)= \underset{{\tl {\bf p}}^2 \le M^2+(\tl p^0)^2+{\bf p}^2}{\int}
    \dd^3 {\bf \tl p} \, \frac{1}{2 \omega}\,
     \left [ c(\omega,{\bf p},\tl p^0, {\bf \tl p})
   +c(-\omega,{\bf p},\tl p^0, {\bf \tl p}) \right ]
\lbl{2.30}
\ee
The initial data for the derivatives of the field with respect to the
time-like coordinates $x^0,{\bf \tl x}$, are
\be
   \frac{\p \phi(0,{\bf x},\tl x^0,{\bf 0})}{\p x^0} =
   \int \dd^3 {\bf p} \, \dd \tl p^0 \, B({\bf p},\tl p^0)
   {\rm e}^{-i({\bf p} {\bf x}+\tl p_0 \tl x^0 )} ,
\lbl{2.30a}
\ee
\be
   \frac{\p \phi(0,{\bf x},\tl x^0,{\bf 0})}{\p {\bf \tl x}} =
   \int \dd^3 {\bf p} \, \dd \tl p^0 \,
   {\bf C}({\bf p},\tl p^0)
   {\rm e}^{-i({\bf p} {\bf x}+\tl p_0 \tl x^0 )} ,
\lbl{2.30b}
\ee
where
\be
   B({\bf p},\tl p^0) = \underset{{\tl {\bf p}}^2 \le M^2+(\tl p^0)^2
   +{\bf p}^2}{\int}
    \dd^3 {\bf \tl p} \, \frac{1}{2} \Bigl [
   c(\omega,{\bf p},\tl p^0,{\bf \tl p})
    + c(-\omega,{\bf p},\tl p^0,{\bf \tl p}) \Bigr ] ,
\lbl{2.30c}
\ee
 \be
   {\bf C}({\bf p},\tl p^0) = \underset{{\tl {\bf p}}^2 \le M^2+(\tl p^0)^2
   +{\bf p}^2}{\int}\dd^3 {\bf \tl p} 
   \, \frac{\bf \tl p}{2 \omega} \Bigl [
   c(\omega,{\bf p},\tl p^0,{\bf \tl p})
    + c(-\omega,{\bf p},\tl p^0,{\bf \tl p}) \Bigr ] ,
\lbl{2.30d}
\ee     
Here ${\bf p},~\tl p^0$ are not restricted, but may have values between
$-\infty,~\infty$, and  $A({\bf p},\tl p_0)$, $B({\bf p},\tl p_0)$,
${\bf C}({\bf p},\tl p_0)$ are arbitrary\footnote{The functions are
arbitrary within the limitation of field theory and quantum mechanics.}
functions of ${\bf p},\tl p^0$. Therefore, $\phi(0,{\bf x},\tl x^0,{\bf 0} )$,
$\p \phi(0,{\bf x},\tl x^0,{\bf 0})/\p x^0$,
$\p \phi(0,{\bf x},\tl x^0,{\bf 0})/\p {\bf \tl x}$ are arbitrary
functions of the space-like coordinates $({\bf x},\tl x^0)$.

From Eq.\,(\ref{2.30}),
(\ref{2.30c}),(\ref{2.30d})
we see that the expansion coefficients $A({\bf p},\tl p_0)$, $B({\bf p},\tl p_0)$,
${\bf C}({\bf p},\tl p_0)$ are the integrals over
the extra time-like momenta ${\tl {\bf p}}\equiv (\tl p^1,\tl p^2,\tl p^3)$,
and that they partially determine the expansion coefficients
$c(\omega,{\bf p},\tl p^0,{\bf \tl p})$ and
$c(-\omega,{\bf p},\tl p^0,{\bf \tl p})$,
and thus give us some, although not complete, information on the behavior
of the field outside the 4-surface on which the initial data are given.
Thus, given the initial data (\ref{2.29a}),(\ref{2.30a}),(\ref{2.30b})
on a space-like 4-surface, the field $\phi(x^\mu,\tl x^\mu)$ is not uniquely
determined. This is in conflict with the requirement that a physical
theory should be deterministic. But is it indeed necessary for a physical
theory to be deterministic? What about quantum theory? In quantum
theory, behavior of a field is not deterministic. Perhaps ultrahyperbolic
field equations point towards the roots of quantum (field) theory.
Since classical field theory, based on an ultrahyperbolic equation, is not
deterministic, it should be supplemented (or replaced) by a more general
theory which takes into account the indeterminism. Such theory is
quantum mechanics. We leave this intriguing point for future investigations.

In particular, if
\bear
  c(\omega,{\bf p},\tl p^0, {\bf \tl p})&=&\delta^3({\bf \tl p}-{\tl {\bf p}}_{\rm c}) 
  a(\omega,{\bf p},\tl p^0) \nonumber\\
c(-\omega,{\bf p},\tl p^0, {\bf \tl p})&=&\delta^3({\bf \tl p}-{\tl {\bf p}}_{\rm c}) 
  a(-\omega,{\bf p},\tl p^0)  ,
\lbl{2.31}
\ear
where ${\bf \tl p}_{\rm c}^2 \le M^2$,
then
\be
    A({\bf p},\tl p_0)= \frac{1}{2 \omega_{{\bf \tl p}_c}}\,
     \left [ a(\omega_{{\bf \tl p}_c}, {\bf p}, \tl p^0)
   +a(-\omega_{{\bf \tl p}_c}, {\bf p}, \tl p^0) \right ]
\lbl{2.32}
\ee
\be
    B({\bf p},\tl p_0)= \frac{1}{2}\,
     \left [ a(\omega_{{\bf \tl p}_c}, {\bf p}, \tl p^0)
   +a(-\omega_{{\bf \tl p}_c}, {\bf p}, \tl p^0) \right ]
\lbl{2.32a}
\ee
\be
    {\bf C}({\bf p},\tl p_0)= \frac{{\bf \tl p}_c}{2 \omega_{{\bf \tl p}_c}}\,
     \left [ a(\omega_{{\bf \tl p}_c}, {\bf p}, \tl p^0)
   +a(-\omega_{{\bf \tl p}_c}, {\bf p}, \tl p^0) \right ] ,
\lbl{2.32b}
\ee
where $\omega_{{\bf \tl p}_c} \equiv
\vert \sqrt{(\tl p^0)^2+{\bf p}^2 -{\bf \tl p}_c^2+M^2} \vert$.
The form (\ref{2.31}) of the expansion coefficients means that the extra time-like
momenta have a fixed definite value  ${\bf \tl p}={\tl {\bf p}}_{\rm c}$. Then we
recover the usual hyperbolic case of one time-like coordinate $x^0$, and
momentum $p^0$. In our theory, we also have an additional extra space
like coordinate, $\tl x^0$, and an additional component of momentum, $\tl p^0$.

If the expansion coefficients are given by Eq.\,(\ref{2.31}),
(\ref{2.32}), and $M^2 - {\bf \tl p}_c^2 >0$,
then the remaining components of momentum, ${\bf p}$, $\tl p^0$ all have
values between $-\infty$ and $+\infty$. The wave packet (\ref{2.28}) has then
the form
\be
   \phi(x^\mu,\tl x^\mu)={\rm e}^{i {\bf \tl p}_{\rm c} {\bf \tl x}}
\, \int \dd^3 {\bf p}\, \dd \tl p^0\,\frac{1}{2 \omega_{{\bf \tl p}_c}} \left [
{\rm e}^{i \omega_{{\bf \tl p}_c} x^0} a(\omega_{{\bf \tl p}_c},\tl p^0,{\bf p}) +
{\rm e}^{-i \omega_{{\bf \tl p}_c} x^0} a(-\omega_{{\bf \tl p}_c},\tl p^0,{\bf p}) 
\right ] {\rm e}^{-i ({\bf p}{\bf x}+\tl p_0 \tl x^0 )}
\lbl{2.35}
\ee
It can be arbitrarily localized on the surface $x^0=0,~{\bf \tl x}=0$ that is
spanned by coordinates $(x^1,x^2,x^3,\tl x^0)$, but it is delocalized
in the time-like directions $(x^0,\tl x^1,\tl x^2,\tl x^3)$ (Fig.\,3a).

\setlength{\unitlength}{.4mm}

\begin{figure}[h!]
\begin{picture}(180,130)(-10,-10)


\newsavebox{\aaa}
\savebox{\aaa}(40, 32)[l]{

\put(25,-52){$a)$}
\put(45.5,0){\vector(1,0){19.5}}
\put(0,0){\line(1,0){14.5}}
\put(0,0){\vector(0,1){65}}
\put(0,0){\vector(-1,-1){35}}

\put(68,1){$\tl x^0$}
\put(-25,69){$x^0,\tl x^1,\tl x^2,\tl x^3$}
\put(-53,-46){$x^1,x^2,x^3$}


\qb(14.5,-25)(17,-31)(30,-32)
\qb(30,-32)(43,-31)(45.5,-25)


\renewcommand{\xscale}{1}
\renewcommand{\xscaley}{-1}
\renewcommand{\yscale}{0.45}
\renewcommand{\yscalex}{0.45}
\put(30,50){\bigcircle{22}}
\put(14.5,50){\line(0,-1){75}}
\put(45.5,50){\line(0,-1){75}}
}

\newsavebox{\bbb}
\savebox{\bbb}(40, 32)[l]{

\put(22,-52){$b)$}
\put(0,0){\vector(1,0){65}}
\put(0,47.5){\vector(0,1){18}}
\put(0,0){\line(0,1){16}}
\put(0,0){\vector(-1,-1){35}}

\put(68,1){$x^1,x^2,x^3,\tl x^0$}
\put(1,67){$x^0$}
\put(-53,-46){$\tl x^1,\tl x^2,\tl x^3$}

\newsavebox{\local}
\savebox{\local}(40, 32)[l]{

\begin{turn}{-90}
\qb(14.5,-25)(17,-31)(30,-32)
\qb(30,-32)(43,-31)(45.5,-25)


\renewcommand{\xscale}{1}
\renewcommand{\xscaley}{-1}
\renewcommand{\yscale}{0.45}
\renewcommand{\yscalex}{0.45}
\put(30,50){\bigcircle{22}}
\put(14.5,50){\line(0,-1){75}}
\put(45.5,50){\line(0,-1){75}}

\end{turn} }

\put(-18,46){\usebox{\local}}

}

\put(60, 35){\usebox{\aaa}}
\put(230, 35){\usebox{\bbb}}

\end{picture}

\caption{ \footnotesize a) A bradyonic field can be localized within a region of
the space-like subspace $(x^1,x^2,x^3,\tl x^0)$, while being delocalized in
the time-like subspace $(x^0,\tl x^1,\tl x^2,\tl x^3)$. The field
is confined within an infinitely long cylinder that extends along an
$(x^0,\tl x^1,\tl x^2,\tl x^3)$-direction. If the bradyon moves with a
nonvanishing speed, the cylinder is suitably inclined.
b) A tachyonic field can be localized within a region of the time-like
subspace $(x^0,\tl x^1,\tl x^2,\tl x^3)$, while being delocalized the
space-like subspace  $(x^1,x^2,x^3,\tl x^0)$. The field is now confined
within a cylinder that is "orthogonal" to the cylinder of Fig.\,3a.
The case of a tachyon with infinite speed is illustrated. For finite tachyon
speed the cylinder is suitably inclined.}
\end{figure}
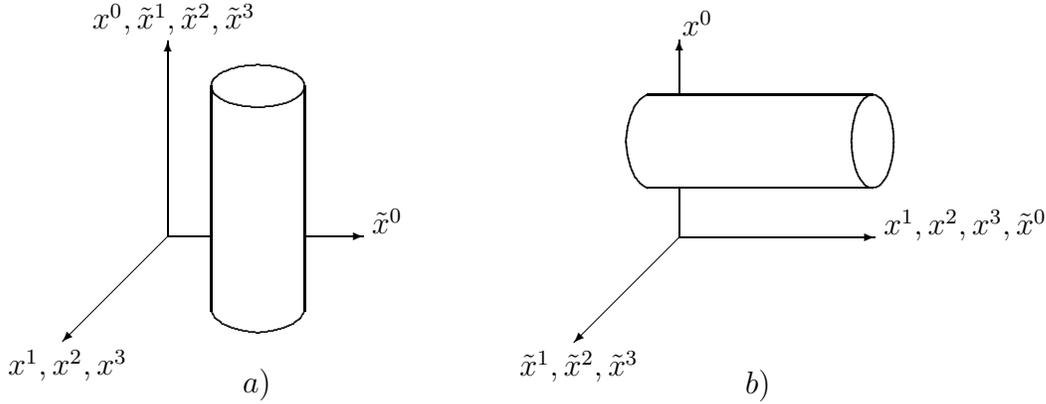

In the case of {\it the tachyonic field}, $M^2<0$, the Klein-Gordon equation
reads
\be
    (\p^\mu \p_\mu - \tl \p^\mu \tl \p_\mu - \tl M^2) \tl \phi(x^\mu,\tl x^\mu)
    =0 ,
\lbl{2.36}
\ee
where $\tl M^2=-M^2 >0$. This equation has the same form as Eq.\,(\ref{2.26a}),
with $x^\mu$ and $\tl x^\mu$ interchanged. Therefore, we can repeat the
same procedure as before. A general solution is
\be
   \tl \phi=\int \dd^4 p \,\dd^4 \tl p \, \tl c(p, \tl p) \,
   {\rm e}^{i(- p_\mu x^\mu +\tl p_\mu \tl x^\mu )}
    \delta (- p^\mu p_\mu + \tl p^\mu \tl p_\mu  -\tl M^2)
\lbl{2.37}
\ee
Now we choose a space-like component, e.g., $\tl p^0$, and integrate it out. Instead of the solution
(\ref{2.28}), we obtain\footnote{
If we choose to integrate out the time-like component $p^0$, then we obtain
   $$\tl \phi(x^\mu,\tl x^\mu)= 
   \int \dd^3 {\bf p}\,\dd \tl p^0\, \dd^3 {\bf \tl p}\, \,
   \frac{1}{2 \omega}\, \left [ {\rm e}^{i (\omega x^0-{\bf p} {\bf x}
   - \tl p_0 \tl x^0 + {\bf \tl p} {\bf \tl x})} \tl c(\omega,{\bf p}, \tl p^0, 
   {\bf \tl p}) + {\rm e}^{i (-\omega \tl x^0-{\bf p} {\bf x}
   -\tl p_0 x^0 +{\bf \tl p} {\bf \tl x})} 
   \tl c(-\omega,{\bf p}, \tl p^0, {\bf \tl p}) \right ] , $$
where
   $ p^0=\pm \omega$, and
    $\omega =\vert \sqrt{{\bf p}^2 +(\tl p^0)^2-{\bf \tl p}^2  -\tl M^2} \vert $.
}
   $$\tl \phi(x^\mu,\tl x^\mu)= \int \dd p^0 \,\dd^3 {\bf p}\, \dd^3 {\bf \tl p}\,
   \frac{1}{2 \tl \omega}\, \left [ {\rm e}^{i (- p_0 x^0 +{\bf p} {\bf x}
  +\tl \omega \tl x^0 - {\bf \tl p} {\bf \tl x})} \tl c(p^0,{\bf p},\tl \omega,  
   {\bf \tl p}) \right .  \hs{2cm}$$
\be
    \hs{4cm} \left . + \, {\rm e}^{i (- p_0 x^0 +{\bf p} {\bf x}
  -\tl \omega \tl x^0 - {\bf \tl p} {\bf \tl x})} \,
   \tl c(p^0,{\bf p},-\tl \omega, {\bf \tl p}) \right ] ,
\lbl{2.38}
\ee
where
\be
    \tl p^0=\pm  \tl \omega~,~~~~~
    \tl \omega =\vert \sqrt{(p^0)^2-{\bf p}^2+{\bf \tl p}^2  +\tl M^2} \vert
\lbl{2.39}
\ee
If, instead of $\tl p^0$, we integrate out from equation (\ref{2.37}) some other
space-like component, e.g., $p^1\equiv p_x$, then we obtain an expression, slightly
different in form, but mathematically equivalent to the expression
(\ref{2.38}):
$$
   \tl \phi(x^\mu,\tl x^\mu) 
   = \int \dd p^0\,\dd p^2\,\dd p^3\,\dd \tl p^0\,\dd {\bf \tl p}\,
   \frac{1}{2 \omega_x} \left [ {\rm e}^{i(-p_0+\omega_x x^1 +p_2 x^2
   +p_3 x^3-\tl p_0 \tl x^0 - {\bf \tl p} {\bf \tl x})}
   \tl c(p^0,\omega_x,p^2,p^3,\tl p^0,{\bf \tl p}) \right . $$
\be
   + \left . {\rm e}^{i(-p_0-\omega_x x^1 +p_2 x^2
   +p_3 x^3-\tl p_0 \tl x^0 - {\bf \tl p} {\bf \tl x})}
   \tl c(p^0,-\omega_x,p^2,p^3,\tl p^0,{\bf \tl p}) \right ] ,
\lbl{2.39a}
\ee
where $p_1 =\pm \omega_x$, $~\omega_x=\Bigl\vert \sqrt{(p^0)^2-(p^2)^2-(p^3)^2
-(\tl p^0)^2 + ({\bf \tl p}^2)^2 + ({\bf \tl p}^3)^2+\tl M^2}\, \Bigl\vert$

A tachyonic field can be arbitrarily localized on the 4-surface $\tl x^0=0$,
${\bf x}=0$ that is spanned by time-like
 coordinates $(x^0,\tl x^1,\tl x^2,\tl x^3)$,
but it cannot be localized in the space-like directions
$(x^1,x^2,x^3,\tl x^0)$ (Fig.\,3b).

For the particular choice of expansion coefficients,
\be
    c(p^0,{\bf p},-\tl \omega, {\bf \tl p})= \delta(p^2)\delta(p^3)\delta(\tl p^0)
     \delta(\tl p^1) a(p^0,\omega_x,\tl p^2,\tl p^3) ,
\lbl{2.39b}
\ee
\be
    c(p^0,{\bf p},-\tl \omega, {\bf \tl p}) = \delta(p^2)\delta(p^3)\delta(\tl p^0)
     \delta(\tl p^1) a(p^0,-\omega_x,\tl p^2,\tl p^3) ,
\lbl{2.39c}
\ee
where $\omega_x=\Bigl\vert\sqrt{(p^0)^2+ ({\bf \tl p}^2)^2 
+({\bf \tl p}^3)^2+\tl M^2}\,\Bigr\vert$,
Eq.\,(\ref{2.39a}) becomes the tachyonic field (\ref{2.12}), considered in Sec.\,2.

Perhaps the fact that neither a bradyonic nor a tachyonic field can be
freely localized on a 7D hypersurface, but there are restrictions on
how they can be localized, should not be considered as problematic at all.
In the literature it is often stated that the differential equations of
physics are a powerful tool, but there is a problem, because we have no
equations
for initial data. The latter data are arbitrary, but it would be desirable
to have a theory for initial data as well. We see that ultrahyperbolic
equations, since they impose certain restriction on initial data,
could be considered as equations that (at least partially) determine the
initial data.

Under a SLT, the Klein-Gordon equation for a tachyonic field transforms into
\be
    (-\p'^\mu \p'_\mu + \tl \p'^\mu \tl \p'_\mu 
    - \tl M^2) \tl \phi'(x'^\mu,\tl x'^\mu)=0 ,
\lbl{2.40}
\ee
which is the Klein-Gordon equation for a bradyonic field
$\phi(x'^\mu,\tl x'^\mu)=\tl \phi'(x'^\mu,\tl x'^\mu)$.

A field that is observed as tachyonic in a reference frame $S$, is observed
as bradyonic in a superluminal reference frame $S'$. In a given reference
frame $S$ we have the bradyonic fields $\phi(x^\mu,\tl x^\mu)$,
satisfying $p^\mu p_\mu -\tl p^\mu \tl p_\mu >0$, and the tachyonic
fields $\tl \phi(x^\mu \tl x^\mu)$, satisfying
$p^\mu p_\mu -\tl p^\mu \tl p_\mu <0$. In a superluminal frame $S'$,
the type of those fields is interchanged.

We have a symmetry between bradyonic and tachyonic fields. None of those
fields is more consistent than the other. Both kinds of fields are
described by the ultrahyperbolic Klein-Gordon equation. We have reduced
the problem of consistent propagating tachyonic fields to the problem of
whether the ultrahyperbolic wave equation can make sense in physics.
In addition, we have a problem of where do the extra dimensions 
$\tl x^\mu$, $\mu=0,1,2,3$, come from and why do we not observe them.
One possibility is just to suppose that our spacetime has not four, but
eight dimensions (or, equivalently, that it is complex), and that the
extra dimensions are not observed, because they are compactified.
Another possibility is to consider
the Clifford space, a 16$D$ manifold whose tangent space at any of its
points is the Clifford algebra of spacetime $M_{1,3}$.

\section{Polyvector-valued coordinates and momenta}

Geometric calculus based on Clifford algebra is a very successful
mathematical tool for description of physics\,\ci{Clifford,Clifford1}. In
addition, it enables formulation of a new
theory\,\ci{CliffSpace}--\ci{PavsicBook} in which
the 4$D$ spacetime is replaced by a more general, 16$D$ space,
called {\it Clifford
space}, $C$. This is a space whose elements are not only points, but
also oriented lines, 2-areas, 3-volumes, and 4-volumes. Such generalization
brings many new theoretical possibilities that are being
explored\,\ci{CliffSpace}--\ci{PavsicBook}.

The squared interval between two points in Minkowski space $M_{1,3}$ is given by
the quadratic form
\be
    Q= (x^\mu-x_0^\mu)\eta_{\mu \nu}(x^\nu-x_0^\nu) .
\lbl{2.40a}
\ee
If we take the square root, we obtain two possible results:
\be
       {\rm (i)}~~~\sqrt{Q} = \Delta s~,  \hs{2.7cm} \hs{2cm}
\lbl{2.41}
\ee
where $\Delta s$ is the {\it scalar distance}, and
\be
    {\rm (ii)} ~~~\sqrt{Q} = \Delta x = (x^\mu-x_0^\mu) \gam_\mu , \hs{2cm}
\lbl{2.42}
\ee
where $\Delta x$ is the {\it vector} that joins the points $x^\mu$ and
$x_0^\mu$. Here the basis vectors $\gam_\mu$ are generators of the
Clifford algebra, $Cl(1,3)$, satisfying
\be
     \gam_\mu \cdot \gam_\nu \equiv \frac{1}{2} (\gam_\mu \gam_\nu
     + \gam_\nu \gam_\mu) = \eta_{\mu \nu} .
\lbl{2.43}
\ee     
The points of $M_{1,3}$ can be described by vectors $x=x^\mu \gam_\mu$,
originating from a chosen fixed point, $x_0^\mu=0$. The 4$D$ space of vectors
$x^\mu \gam_\mu$ is a subspace of the 16$D$ Clifford algebra $Cl(1,3)$,
whose elements are {\it polyvectors}  
\be
    X = \sigma {\bf 1} + x^\mu \gam_\mu + x^{\mu \nu} \gam_\mu \wg \gam_\nu
    +\tl x^\mu \gam_5 \gam_\mu + \tl \sigma \gam_5 \equiv x^M \gam_M .
\lbl{2.44}
\ee
Here, $\gam_5 = \gam_0 \gam_1 \gam_2 \gam_3$, and `$\wg$' the wedge
product, e.g., $\gam_\mu \wg \gam_\nu =$ $(1/2) (\gam_\mu \gam_\nu -
\gam_\nu \gam_\mu)$. Polyvectors, or $r$-vectors,
describe oriented $r$-volumes (called also $r$-areas), $r=0,1,2,3,4$.
The idea is\,\ci{Castro}--\ci{PavsicArena} that $r$-volumes are associated with
extended events living in spacetime $M_{1,3}$. Thus, e.g., 2-volumes that
are described by $x^{\mu \nu} \gam_\mu \wg \gam_\nu$ are associated with
the class of those physical objects, whose representative is a closed
{\it instantonic} string, or alternatively, and open {\it instantonic}
2-brane, depending on whether the scalar $\sigma$ is zero, or different
from zero. Namely, the scalar $\sigma$ can be defined in
terms of the length, area, 3-volume, and 4-volume of, respectively,
the open 1-brane, 2-brane, 3-brane and 4-brane. For instance,
for an open instantonic string (i.e., 1-brane), the scalar is defined as
\be
      \sigma_1 = \frac{{\cal A}_1}{L_P}=\frac{1}{L_P} \int \dd \xi
      \left ( \frac{\p X^\mu}{\p \xi}\frac{\p X^\nu}{\p \xi} \eta_{\mu \nu}
      \right )^{1/2} ,
\lbl{2.45}
\ee
where ${\cal A}_1$ is the string length, and $\xi$ an arbitrary real parameter
denoting the points on the string.
In general, for an open instantonic $r_{\sigma}$-brane, we have
\be
      \sigma_{r_\sigma} = \frac{{\cal A}_{r_\sigma}}{(L_P)^{r_\sigma}}=
      \frac{1}{(L_P)^{r_\sigma}} \int \dd \xi^1 \dd \xi^2 ...\dd \xi^{r_\sigma}
    \sqrt{{\rm det}\, \frac{\p X^\mu}{\p \xi^a}
    \frac{\p X^\nu}{\p \xi^b} \eta_{\mu \nu} } ,
\lbl{2.46}
\ee
where ${\cal A}_{r_\sigma}$ is the $r_\sigma$-dimensional area/volume of the
brane, and $\xi^a$ or $\xi^b$, $a,b = 1,2,..., r_\sigma$, the parameters
(coordinates) denoting the points of the brane.
Here, $L_P$ is a fundamental length, for instance the Planck length,

For a generic object, that is a conglomerate of instantonic $r_\sigma$-branes,
for various values of $r_\sigma=1,2,3,4$, the scalar is defined as the
sum
\be
    \sigma = \sum_{r_\sigma=1}^4 \frac{{\cal A}_{r_\sigma}}{(L_P)^{r_\sigma}} .
\lbl{2.46a}
\ee
The scalar $\sigma$ thus determines to what extent the object, described
by a polyvector (\ref{2.44}), and envisaged as a conglomerate of instantonic
$r$-branes, $r=0,1,2,3,4$, contains closed $(r_\sigma -1)$-branes and
open $r_\sigma$-branes, $r_\sigma=1,2,3,4$. If $\sigma=0$, then only closed
$(r_\sigma -1)$-branes are present.

Our objects are {\it instantonic} $r$-branes, which mean that they are
localized in spacetime.\footnote{The usual $p$-branes are localized
in space, but they are infinitely extended into a time-like direction;
therefore, they are $(p+1)$-dimensional worldsheets in spacetime.}
They are a generalization of the concept of an `{\it event}' to which there
corresponds a point in spacetime. Instead of a point, we have now in general
a set of $r$-volumes, $x^{\mu_1 ...\mu_r}$, $r=0,1,2,3,4$, that describe an
{\it extended event} in spacetime.\footnote{Polyvector coordinates
$x^{\mu_1 ...\mu_r}$ describe
a class of $r$-branes, all having the same $x^{\mu_1 ...\mu_r}$. See
Ref.\,\ci{PavsicArena,PavsicMaxwellBrane}.}
The space of the extended
events is called Clifford space, $C$. It is manifold whose tangent space
at any of its points is a Clifford algebra $Cl(1,3)$. If $C$ is a flat
space, then it is isomorphic to the Clifford algebra with elements
$X= x^{\mu_1 ...\mu_r} \gam_{\mu_1 ...\mu_r}\equiv x^M \gam_M$
(Eq.\,(\ref{2.44})).

Instead of the relativity in spacetime, we have now the relativity in Clifford
space\,\ci{CliffSpace}-\ci{PavsicBook}. The line element is
\be
   \dd S^2 = \dd x^M \dd x^N G_{MN} ,
\lbl{2.47}
\ee
where the metric is defined in terms of the scalar product between two
basis elements:
\be
     G_{MN} = \gam_M^\ddg * \gam_N = \langle \gam_M^\ddg \gam_N \rangle_0 .
\lbl{2.48}
\ee
Here $\langle~\rangle_0$ denotes the scalar part, and `$\,^\ddg\,$' reversion,
so that $\gam_M^\ddg \equiv (\gam_{\mu_1} \gam_{\mu_2}...\gam_{\mu_r})^\ddg=
\gam_{\mu_r}\gam_{\mu_r -1}...\gam_{\mu_2} \gam_{\mu_1}$. From the
definition (\ref{2.48}) we obtain the following explicit form for the
line element:
$$
   \dd S^2 = \dd \sigma^2+(\dd x^0)^2-(\dd x^1)^2-(\dd x^2)^2-(\dd x^3)^2
     -(\dd x^{01})^2-(\dd x^{02})^2 -(\dd x^{03})^2  \hs{9mm}$$
\be      
    \hs{12mm} +(\dd x^{12})^2+(\dd x^{13})^2 +(\dd x^{23})^2
     -(\dd \tl x^0)^2+(\dd \tl x^1)^2+(\dd \tl x^2)^2+(\dd \tl x^3)^2
     - \dd \tl \sigma^2 .
\lbl{2.49}
\ee
We see that the signature is $(8,8)$.

Coordinates $x^M$ denote a point in Clifford space, $C$. A point in $C$ is
associated with an {\it event} in $C$, which in turn corresponds to an
extended event in spacetime $M_{1,3}$. Let us now consider a {\it family
of events}, described by a worldline $x^M = X^M (\zeta)$ in $C$, where
$\zeta$ is an arbitrary monotonically increasing parameter along the worldline.
We assume that such a wordline, if associated with a physical object
living in $C$, satisfies the ``minimal" length action
\be
    I[X^M] = M \int \dd \zeta ({\dot X}^M \dot X^N G_{MN} )^{1/2} ,
\lbl{2.50}
\ee
where ${\dot X}^M \equiv \dd X^M/\dd \zeta$.
This is an action for a point particle in $C$. In spacetime $M_{1,3}$, such
particle is extended and traces a ``thick" worldline.

The action (\ref{2.50}) is invariant under reparametrizations of $\zeta$.
As a consequence, the canonical momenta $P_M = \p L/\p \dot X^M =
M \dot X^M/(\dot X^N \dot X_N)^{1/2}$ satisfy the constraint
\be
    G^{MN} P_M P_N - M^2 = 0 .
\lbl{2.51}
\ee

\subsection{Generalized Klein-Gordon equation}

Upon quantization, the constraint (\ref{2.51}) becomes the ultrahyperbolic
Klein-Gordon equation
\be
     \left ( G^{MN} \p_M \p_N + M^2 \right ) \phi (x^M) = 0 ,
\lbl{2.52}
\ee
or explicitly,
\be
   \left ( \frac{\p^2}{\p \sigma^2}+\p^\mu \p_\mu - \tl \p^\mu \tl \p_\mu
   +\p^{\mu \nu} \p_{\mu \nu} - \frac{\p^2}{\p \tl \sigma^2} + M^2
   \right ) \phi(\sigma,x^\mu,x^{\mu \nu},\tl x^\mu, \tl \sigma) = 0 ,
\lbl{2.53}
\ee
which is of the same form as Eq.\,(\ref{2.26a}), except that now the
space is 16-dimensional, with signature $(8,8)$. The coordinates are
$x^M = (\sigma,x^\mu,x^{\mu \nu},\tl x^\mu, \tl \sigma)$, and the metric
is $G_{MN}$, given by Eq.\,(\ref{2.49}).

If we introduce the new coordinates
\be
   \tau= \frac{1}{\sqrt{2}} (\tl \sigma - \sigma) ~,~~~~~
   \lambda = \frac{1}{\sqrt{2}} (\tl \sigma + \sigma) ~,
\lbl{2.54}   
\ee
the Klein-Gordon equation becomes
\be
    \left ( -2 \frac{\p^2}{\p \tau \p \lambda}+\p^\mu \p_\mu
    -\tl \p^\mu \tl \p_\mu
     + \p^{\mu \nu}\p_{\mu \nu} +M^2 \right )
      \psi (\tau,\lambda,x^\mu,\tl x^\mu,x^{\mu \nu})=0 .
\lbl{2.55}
\ee
We will write this compactly as
\be
  (-2 \p_\tau \p_\lambda + G^{\bar \mu \bar \nu}\p_{\bar \mu}\p_{\bar \nu}
   +M^2) \psi(\tau,\lambda,x^{\bar \mu}) =0 ,
\lbl{2.56}
\ee
where $x^{\bar \mu}=(x^\mu,\tl x^\mu,x^{\mu \nu})$, and $G^{\bar \mu \bar \nu}$
the metric with signature $(7,7)$.

A general solution of Eq.\,(\ref{2.56}) is
\be
   \psi = \int \dd p_\tau \, \dd p_\lambda \dd^{14} \bar p \,
   \, c (p_\tau,p_\lambda,p_{\bar \mu})\, {\rm e}^{i(p_\tau \tau + p_\lambda \lambda 
 +p_{\bar \mu} x^{\bar \mu})}  
\delta (2 p_\tau p_\lambda -G^{\bar \mu \bar \nu}p_{\bar \mu}p_{\bar \nu} +M^2) .
\lbl{2.57}
\ee
Performing the integration over $p_\tau$, we have
\be
   \psi = \int \dd p_\lambda \, \dd^{14} \bar p \, 
    a(\omega,p_\lambda,p_{\bar \mu})
  \,{\rm e}^{i \omega \tau}{\rm e}^{i p_{\bar \mu} x^{\bar \mu}}
   \, {\rm e}^{i p_\lambda \lambda},
\lbl{2.58}
\ee
where
\be
    \omega = \frac{1}{2 p_\lambda} 
    \left ( G^{\bar \mu \bar \nu}p_{\bar \mu}p_{\bar \nu} -M^2 \right ) .
\lbl{2.59}
\ee
and
\be
    a(\omega,p_\lambda,p_{\bar \mu})\equiv \frac{1}{2 p_\lambda}
   \,c(\omega,p_\lambda,p_{\bar \mu}) .
\lbl{2.59a}
\ee
All values of $p_\lambda,p_{\bar \mu}$ between $-\infty$ and $+\infty$ are
allowed. The initial data, given at $\tau=0$, are
\bear
   &&\psi(\tau=0,\lambda,x^{\bar \mu}) = \int \dd p_\lambda \, \dd^{14} \bar p \, 
   \, a(\omega,p_\lambda,p_{\bar \mu})
   \, {\rm e}^{i p_{\bar \mu} x^{\bar \mu}} \, {\rm e}^{i p_\lambda \lambda},
\lbl{2.60} \\
   &&\dot \psi(\tau=0,\lambda,x^{\bar \mu}) = \int \dd p_\lambda \, \dd^{14} \bar p \, 
   \, i \omega \,a(\omega,p_\lambda,p_{\bar \mu})
   \, {\rm e}^{i p_{\bar \mu} x^{\bar \mu}} \, {\rm e}^{i p_\lambda \lambda}.
\lbl{2.61}
\ear
The latter equations tell us that $\psi(\tau=0,\lambda,x^{\bar \mu})$
and $\dot \psi(\tau=0,\lambda,x^{\bar \mu})$ are expanded in terms
of the complete set of functions ${\rm exp}[i (p_\lambda \lambda +
p_{\bar \mu} x^{\bar \mu})]$, and that there is no restriction on initial data,
except that $\psi$ must be normalizable according to
$\int \dd \lambda \, \dd^{14} \bar x \, \vert \psi \vert^2 = 1$. By the Fourier
transform, the expansion coefficients $a(\omega,p_\lambda,p_{\bar \mu})$
can be expressed in terms of the initial
data. As a consequence, $\psi (\tau,\lambda, x^{\bar \mu})$ is uniquely
determined at any $\tau>0$. The Cauchy problem is thus well defined\footnote{
That the Cauchy problem for an ultrahyperbolic partial differential equation is
well defined for initial data, taken at a light-like hypersurface, is
a known result\,\ci{CauchyLightLike}.}.

   If we take
\be
    a(\omega,p_\lambda,p_{\bar \mu})
    =\delta (p_\lambda - \Lambda) A(p_{\bar \mu}) ,
\lbl{2.62}
\ee
then
\be
    \psi(\tau,\lambda, x^{\bar \mu})=\int \dd^{14} \bar p\, A(p_{\bar \mu})
   {\rm e}^{i \omega \tau} 
    {\rm e}^{i p_{\bar \mu} x^{\bar \mu}} \, {\rm e}^{i \Lambda \lambda}
    \equiv \varphi(\tau,x^{\bar \mu}) {\rm e}^{i \Lambda \lambda} ,
\lbl{2.63}
\ee
where
\be
    \omega = \frac{1}{2 \Lambda} 
    \left ( G^{\bar \mu \bar \nu}p_{\bar \mu}p_{\bar \nu} -M^2 \right ) .
\lbl{2.64}
\ee
This is a solution of Eq.\,(\ref{2.56}) with the definite value 
$p_\lambda=\Lambda$, of the momentum operator $\hat p_\lambda = -i \/\p_\lambda$.
By plugging Eq.\,(\ref{2.63}) into Eq.\,(\ref{2.56}), we obtain the wave
equation for $\varphi(\tau,x^{\bar \mu})$:
\be
    i \, \frac{\p \varphi(\tau,x^{\bar \mu})}{\p \tau}=\frac{1}{2 \Lambda}
    (G^{\bar \mu \bar \nu}\p_{\bar \mu}\p_{\bar \nu} +M^2) 
    \varphi(\tau,x^{\bar \mu}) .
\lbl{2.65}
\ee
This is the generalized\footnote{The usual Stueckelberg
equation\ci{Stueckelberg,Stueckelberg1} involves
4 spacetime coordinates $x^\mu$, $\mu=0,1,2,3$, whereas the generalized
equation (\ref{2.65}) involves 14 coordinates $x^{\bar \mu}=
(x^\mu,\tl x^\mu,x^{\mu \nu})$.}
 {\it Stueckleberg equation} for the wave function
$\varphi(\tau,x^{\bar \mu})$. It is like the Schr\"odinger equation, the
evolution parameter being $\tau$, and the wave function being normalized
according to $\int \dd^{14} {\bar x} \, |\varphi(\tau,x^{\bar \mu})|^2 = 1$.

From the wave equation (\ref{2.65}) and its solution (\ref{2.63}) we see that
$M^2$ gives a constant phase factor ${\rm exp} [-i(1/2 \Lambda) M^2 \tau ]$, and
has no other role. It makes not much difference if $M^2>0$ or $M^2<0$. The
solution of Eq.\,(\ref{2.65}) is thus not sensitive on whether the field is
bradyonic or tachonic in the 16$D$ space $C$.

On the other hand, it is well-known that the
Stueckelberg\,\ci{Stueckelberg,Stueckelberg1}
theory admits superluminal
motion in 4$D$ spacetime $M_{1,3}$. All 14 momenta $p_{\bar \mu}
=(p_\mu,\tl p_{\bar \mu},p_{\mu \nu})$ occurring in the solution (\ref{2.63}) are
independent, and their quadratic form 
$G^{\bar \mu \bar \nu}p_{\bar \mu}p_{\bar \nu}\equiv \bar p^2$ can be either
positive or negative. The same is true for the quadratic form of the spacetime
momenta, $\eta^{\mu \nu} p_\mu p_\nu \equiv p^2$. It can be $p^2>0$, in which
case the field behaves as subluminal in $M_{1,3}$. But it also can be
$p^2<0$, and then the field is tachyonic in $M_{1,3}$. In the presence of
suitable interactions\footnote{An example of an interaction that can accelerate
a classical Stueckelberg point particle from subluminal to superluminal speeds
is considered in Ref.\ci{PavsicBook}. See also
Refs.\,\ci{Stueckelberg, Stueckelberg1}.},
there can be smooth transition from the subluminal to the superluminal case.

\subsection{Generalized Dirac equation}

In Refs.\,\ci{CliffordPossibilities,CliffordPossibilities1} the following generalization of the Dirac equation
for a polyvector valued wave funtion was proposed:
\be
    \gam^M \p_M \Phi = 0 ,
\lbl{a1}
\ee
where $\Phi = \phi^A \gam_A =\phi^{\tl A} \xi_{\tl A}$ is expanded in terms
of the Clifford algebra basis, $\gam_A = (1,\gam_a, \gam_a \wg \gam_b,...)$,
or in terms pf the generalized spinor basis
$\xi_{\tl A} \equiv \xi_{\alpha i}$. Here $\alpha=1,2,3,4$ is the spinor
index, and $i=1,2,3,4$ the index denoting four minimal left ideals
of $Cl(1,3)$, a tangent space of the Clifford space, $C$. The components
$\phi^{\tl A} = \phi^{\tl A} (x^M)$ are complex valued fields on $C$.

Explicitly, Eq.\,(\ref{a1}) reads
\be
   \left ( \frac{\p }{\p \sigma} + \gam^\mu \p_\mu + \gam^\mu \gam^\nu
   \p_{\mu \nu} + \gam^5 \gam^\mu {\tl \p}_\mu + \gam^5 \frac{\p}{\p \tl \sigma}
    \right ) \Phi = 0.
\lbl{a2}
\ee

Let us now consider two possible kinds of solutions, each satisfying a different
Ansatz.

{\it Ansatz I}
\be
    \Phi= {\rm e}^{i m \sigma} {\rm e}^{i \tl m \tl \sigma}
    \Psi(x^\mu,x^{\mu \nu},{\tl x}^\mu)
\lbl{a3}
\ee
\be
    \frac{\p \Phi}{\p \sigma} = i m \Phi~,~~~~~~
     \frac{\p \Phi}{\p \tl \sigma} = i \tl m \Phi
\lbl{a4}
\ee
Then Eq.\,(\ref{a2}) becomes
\be
   \left ( \gam^\mu \p_\mu + \gam^\mu \gam^\nu
   \p_{\mu \nu} + \gam^5 \gam^\mu {\tl \p}_\mu + i m + i \tl m \gam^5
    \right ) \Psi = 0.
\lbl{a5}
\ee

{\it Ansatz II}
\be
    \Phi= {\rm e}^{- m \sigma} {\rm e}^{- \tl m \tl \sigma}
    \Psi(x^\mu,x^{\mu \nu},{\tl x}^\mu)
\lbl{a6}
\ee
\be
    \frac{\p \Phi}{\p \sigma} = -m \Phi~,~~~~~~
     \frac{\p \Phi}{\p \tl \sigma} = - \tl m \Phi
\lbl{a6a}
\ee
Then Eq.\,(\ref{a2}) becomes
\be
   \left ( \gam^\mu \p_\mu + \gam^\mu \gam^\nu
   \p_{\mu \nu} + \gam^5 \gam^\mu {\tl \p}_\mu - m - \tl m \gam^5
    \right ) \Psi = 0.
\lbl{a7}
\ee

A special case of the latter equation is
\be
      (\gam^\mu \p_\mu - \tl m \gam^5 ) \Psi ,
\lbl{a8}
\ee
which holds when $\Psi$ does not depend on $x^{\mu \nu}$, $\tl x^\mu$, and
$\sigma$. Multiplying (\ref{a8}) from the left by $\gam^5$, we obtain
\be
    (\gam^5 \gam^\mu \p_\mu + \tl m ) \Psi  = 0 .
\lbl{a9}
\ee
Further we have
\be
   ((\gam^5 \gam^\mu)^\ddg \p_\mu + \tl m )
    (\gam^5 \gam^\mu \p_\mu + \tl m ) \Psi 
    =( -\eta^{\mu \nu} \p_\mu \p_\nu + \tl m^2 ) \Psi = 0.
\lbl{a10}
\ee
This is the ``tachyonic'' Klein-Gordon equation\footnote{
As discussed in Sec.\,2, this is not the true tachyonic Klein-Gordon
equation, because it cannot be obtained from a bradyonic equation by a SLT.
It is a part of the complete equation (\ref{a16}), which is either tachyonic
or bradyonic, depending on the sign of $m^2-\tl m^2$.},
satisfying the 4-momentum
constraint $p^\mu p_\mu = - \tl m^2$. The same constraint holds
for Eq.\,(\ref{a9}), which is therefore the ``tachyonic'' Dirac equation.
Such equation\footnote{ We use here the definition $\gam^5=
\gam^0 \gam^1 \gam^2 \gam^3$, satisfying $(\gam^5)^2 = -1$, $(\gam^5)^\ddg =
\gam^5$, that is customary in the literature on Clifford algebras.
Chodos et al.\,\ci{Chodos} use the definition $\gam^5=i\gam^0 \gam^1 \gam^2 \gam^3$,
satisfying $(\gam^5)^2 = 1$, $(\gam^5)^\ddg = -\gam^5$, $i^\ddg = -1$.}
was proposed by Chodos et al.\,\ci{Chodos}. It is a special case of
the generalized Dirac equation (\ref{a1}). Solutions of Eq.\,(\ref{a9})
and their localizability were investigated in Ref.\,\ci{Jentschura3}.
A different approach to a tachyonic Dirac equation is considered in
Ref.\,\ci{Chang}. The imaginary mass Dirac equation and its quantization
has been investigated by Jentschura\,\ci{Jentschura4} who found that
right handed states acquire a negative Fock-space norm, which could explain
the absence of right-handed neutrinos.

If we start from the original equation (\ref{a1}), multiply it by
${\gam^{M}}^\ddg$ and take the scalar part, we have
\be
   \langle {\gam^M}^\ddg \gam^N \rangle_0 \,\p_M \p_N \Phi = 0 ~, ~~~~
   i.e.,~~~G^{MN}\p_M \p_N \Phi = 0,
\lbl{a13}
\ee
or explicitly,
\be
  \left ( \frac{\p^2}{\p \sigma^2}+ \eta^{\mu \nu} \p_\mu \p_\nu + \p^{\mu \nu}
   \p_{\mu \nu} - \eta^{\mu \nu} \tl \p_\mu \tl \p_\nu
    - \frac{\p^2}{\p \tl \sigma^2} \right ) \Phi = 0,
\lbl{a14}
\ee
which is Eq.\,(\ref{2.53}) for $M=0$.

Eq.\,(\ref{a14}), in the case of Ansatz I becomes
\be
  \left ( \p^\mu \p_\nu + \p^{\mu \nu} \p_{\mu \nu}
   - \tl \p^\mu \tl \p_\nu -m^2 + \tl m^2 \right ) \Psi = 0,
\lbl{a15}
\ee
and in the case of Ansatz II, it becomes
\be
  \left ( \p^\mu \p_\nu + \p^{\mu \nu} \p_{\mu \nu}
   - \tl \p^\mu \tl \p_\nu +m^2 - \tl m^2 \right ) \Psi = 0,
\lbl{a16}
\ee
Both equations, (\ref{a15}),(\ref{a16}), are either bradyonic or tachyonic,
depending on the sign of $m^2-\tl m^2$.

Besides Ansatz I and Ansatz II, we can also consider

{\it Ansatz III}
\be
    \Phi= {\rm e}^{- m \sigma} {\rm e}^{i\tl m \tl \sigma}
    \Psi(x^\mu,x^{\mu \nu},{\tl x}^\mu) ,
\lbl{a17}
\ee
with
\be
   \left ( \gam^\mu \p_\mu + \gam^\mu \gam^\nu
   \p_{\mu \nu} + \gam^5 \gam^\mu {\tl \p}_\mu - m + i \tl m \gam^5
    \right ) \Psi = 0,
\lbl{a18}
\ee
\be
 \left ( \p^\mu \p_\nu + \p^{\mu \nu} \p_{\mu \nu}
   - \tl \p^\mu \tl \p_\mu + m^2 + \tl m^2 \right ) \Psi = 0, 
\lbl{a19}
\ee
and

{\it Ansatz IV}
\be
    \Phi= {\rm e}^{i m \sigma} {\rm e}^{- \tl m \tl \sigma}
    \Psi(x^\mu,x^{\mu \nu},{\tl x}^\mu) ,
\lbl{a20}
\ee
with
\be
   \left ( \gam^\mu \p_\mu + \gam^\mu \gam^\nu
   \p_{\mu \nu} + \gam^5 \gam^\mu {\tl \p}_\mu +i m - \tl m \gam^5
    \right ) \Psi = 0.
\lbl{a21}
\ee
\be
  \left ( \p^\mu \p_\nu + \p^{\mu \nu} \p_{\mu \nu}
   - \tl \p^\mu \tl \p_\mu - m^2 - \tl m^2 \right ) \Psi = 0,
\lbl{a22}
\ee
For real $m$, $\tl m$, Eqs.\,(\ref{a18}),(\ref{a19}) are bradyonic,
whereas Eqs.\,(\ref{a21}),(\ref{a22}) are tachyonic.

If $\p_{\mu \nu} \Psi =0,~\tl \p_\mu \Psi=0$, then Eqs.\,(\ref{a18}),
(\ref{a21}) become
\be
   \left ( \gam^\mu \p_\mu - m + i \tl m \gam^5 \right ) \Psi = 0,
\lbl{a23}
\ee
\be
   \left ( \gam^\mu \p_\mu +i m - \tl m \gam^5 \right ) \Psi = 0,
\lbl{a24}
\ee
Equations of the latter kind were considered in Refs.\,\ci{Jentschura2}.

Usage of the solutions with  ${\rm Exp} [-m \sigma]$
and ${\rm Exp} [-\tl m \tl \sigma]$
makes sense only, if the boundary conditions in the space $(\sigma,\tl \sigma)$
are such that they prevent the escape of functions into infinity.

Our choice of basis,
\be
 \gam_M = (1,\gam_\mu,\gam_\mu \wg\gam_\mu, \gam^5 \gam_\mu, \gam^5 )
\lbl{b1}
\ee
leads to Eqs.\,(\ref{a5}),(\ref{a7}),(\ref{a18}),(\ref{a21}),
corresponding to the choices Eqs.\,(\ref{a3}),(\ref{a6})(\ref{a17}),(\ref{a20}),
respectively. We have seen that Eqs.\,(\ref{a7}) and (\ref{a21}) contain the
so called ``tachyonic'' equation (\ref{a8}) of Chodos et al.\,\ci{Chodos}.

Eqs.\,(\ref{a7}) and (\ref{a18}) contain, as a special case, the following
equation:
\be
   (\gam^\mu \p_\mu - m )\Psi = 0 ,
\lbl{b2}
\ee
which, when squared according to (\ref{a13}), gives the bradyonic Klein-Gordon
equation. But Eq.\,(\ref{b2}) is not quite the usual Dirac equation,
because an $i$ is missing. It is straightforward to verify that an
alternative choice of basis, namely
\be
 \gam_M = (-i,\gam_\mu,\gam_\mu \wg\gam_\mu, \gam^5 \gam_\mu, \gam^5 ) ,
\lbl{b3}
\ee  
gives the ordinary Dirac equation,
\be
   (i \gam^\mu \p_\mu - m )\Psi = 0 ,
\lbl{b4}
\ee
which when squared according to (\ref{a13}), and taking into account
$i^\ddg = -i$, also gives the bradyonic Klein-Gordon equation.

With the basis (\ref{b3}), instead of Eqs.\,(\ref{a23}),(\ref{a24}), we have
\be
   \left (\gam^\mu \p_\mu +i m +i \tl m \gam^5 \right ) \Psi = 0,
\lbl{b5}
\ee
\be
   \left (\gam^\mu \p_\mu + m - \tl m \gam^5 \right ) \Psi = 0,
\lbl{b6}
\ee
which, when multiplied by $i$, are the generalized bradyonic
and tachyonic Dirac equations studied in Ref.\,\ci{Jentschura2}.

To avoid problems with localizability of the field $\Psi$, let us again
consider the light cone like coordinates (\ref{2.54}). In such
coordinates the generalized Dirac equation (\ref{a2}) reads
\be
   -\frac{1}{\sqrt{2}}(1-\gam^5) \p_\tau \Psi 
   + \frac{1}{\sqrt{2}}(1+\gam^5) \p_\lambda \Psi
   + \gam^{\bar \mu} \p_{\bar \mu} \Psi = 0 .
\lbl{a26}
\ee
Using
\be
    \frac{1}{\sqrt{2}}(1+\gam^5)\frac{1}{\sqrt{2}}(1-\gam^5)=1 ,
\lbl{a27}
\ee
\be
    \frac{1}{\sqrt{2}}(1+\gam^5)\frac{1}{\sqrt{2}}(1+\gam^5)=\gam^5 ,
\lbl{a28}
\ee
Eq.\,(\ref{a26}) can be rewritten in the form
\be
    - \p_\tau \Psi + \gam^5 \p_\lambda \Psi + \frac{1}{\sqrt{2}}(1+\gam^5)
       \gam^{\bar \mu} \p_{\bar \mu} \Psi = 0 .
\lbl{a29}
\ee
Eqs.\,(\ref{a26}),(\ref{a29}) satisfy the equation
\be
    (-2 \p_\tau \p_\lambda + \p^{\bar \mu} \p_{\bar \mu}) \Psi =0,
\lbl{a26a}
\ee
which is Eq.\,(\ref{2.56}) with $M=0$.

If $\Psi$ is an eigenstate of the operator $-i\p_\lambda$, so that
$-i\p_\lambda \Psi = \Lambda \Psi$, then Eq.\,(\ref{a29}) becomes
\be
   \left ( - \p_\tau + i \Lambda \gam^5 + \frac{1}{\sqrt{2}}(1+\gam^5)
       \gam^{\bar \mu} \p_{\bar \mu} \right ) \Psi = 0 .
\lbl{a30}
\ee       
Here $\tau$ is the Stueckelberg evolution parameter. In principle,
solutions of Eq.\,(\ref{a30}) should be found in analogous way
as solutions of Eq.\,(\ref{2.56}), by taking into
account the dispersion relation  $-2 p_\tau p_\lambda + p^{\bar \mu}
p_{\bar \mu} = 0$ that comes from Eq.\,(\ref{a26a}).
However, a difference is that now $\Psi$ is a generalized spinor field.
It is localized on a 15-dimensional hypersurface $\tau=constant$, spanned
by coordinates $\lambda, x^{\bar \mu}$, where $x^{\bar \mu} 
= (x^\mu,\tl x^\mu,x^{\mu \nu})$.

If $\Psi$ is also an eigenstate of $-i \p_\tau$, so that
$-i\p_\tau \Psi = K \Psi$, then we obtain the equation
 \be
   \left ( - i K + i \Lambda \gam^5 + \frac{1}{\sqrt{2}}(1+\gam^5)
       \gam^{\bar \mu} \p_{\bar \mu} \right ) \Psi = 0 ,
\lbl{a31}
\ee    
or, equivalently (c.f., Eq.\,(\ref{a26})),
\be
\left ( 
   -\frac{1}{\sqrt{2}}(1-\gam^5) i K  + \frac{1}{\sqrt{2}}(1+\gam^5) i \Lambda 
   + \gam^{\bar \mu} \p_{\bar \mu} \right ) \Psi = 0 .
\lbl{a32}
\ee
If in the latter equation we rearrange the terms and set
$m=(1/\sqrt{2})(\Lambda - K)$, 
$~~\tl m = (1/\sqrt{2})(\Lambda + K)$,
then we obtain Eq.\,(\ref{a5}), which confirms the consistency of the
procedure. If the eigenvalues $K$ and $\Lambda$ are real or imaginary,
then we reproduce Eqs.\,(\ref{a7}),(\ref{a18}) and (\ref{a21}).
Special cases of the last two equations are (\ref{a23}),(\ref{a24}), that
in terms of the alternative Clifford algebra basis (\ref{b3}), can be written
in the form (\ref{b5}),(\ref{b6}). Solutions of the latter equations and their
physical implications have been considered by Jentschura
and Wundt\,\ci{Jentschura2}.

\section{Discussion}

We have reduced the problem of consistent propagating tachyonic field to the
problem of whether the Stueckelberg field makes sense in physics. We
assumed that physics has to be formulated in the 16$D$ Clifford
space $C$, a manifold whose tangent space at any point is Clifford algebra.
Since the signature of $C$ is $(8,8)$, we arrived upon quantization at
the ultrahyperbolic Klein-Gordon or Dirac equation. The Cauchy problem for such
equation is not well posed, unless we take initial data on
a light-like hypersurface. Then, after using the light cone coordinates,
such that the evolution parameter $\tau$ is a superposition of the scalar and
pseudoscalar coordinate, the Klein-Gordon and the Dirac equation become
the corresponding  generalized
Stueckelberg equations. A different choice of light cone
coordinates, involving the time-like coordinate $x^0$, although mathematically
admissible, would be physically problematic, because we do not live on a light
cone of $M_{1,3}$.
A virtue of the Clifford space is that it contains extra time-like and
space-like dimensions, besides the ordinary $x^\mu$, $\mu=0,1,2,3$.
Therefore, in $C$ there exist light-like hypersurfaces that do not
involve $x^\mu$, and one can pose the initial value problem on one of such
hypersurfaces. However, the evolution parameter is then not $x^0$, but
some other parameter associated with a family of hypersurfaces. In our case,
such a parameter is $\tau$, the light cone coordinate (\ref{2.54}).
According to this theory, we live on a light-like hypersurface
in Clifford space, and experience an evolution whose
parameter is $\tau$. The coordinate $x^0$ has now the same status as
the spatial coordinates $x^i$, $i=1,2,3$. The
$x^i$ of a physical object change with evolution, and so does $x^0$.
The wave packet can now be localized not only around $x^i$, but also around
$x^0$, i.e., around a point (``event") ${\cal P}$ in $M_{1,3}$. And with
increasing
$\tau$, the point ${\cal P} (\tau)$ moves in $M_{1,3}$. Since the latter space
is a subspace of the 16$D$ Clifford space $C$, we have in fact a wave packet
localized around a point ${\cal E}$, determined by the
coordinates $x^{\bar \mu}=(x^{\mu},\tl x^\mu,x^{\mu \nu})$ of the 14$D$
space $M_{7,7}\subset C$. With increasing $\tau$, the point ${\cal E}(\tau)$ moves in
$M_{7,7}$. The fact that we, observers, do not experience all of the
spacetime $M_{1,3}$ or all of $M_{7,7}$ at once, is due to the localization
of the wave packet. A further discussion of this topics can be found
in Refs.\,\ci{PavsicEvolution,PavsicBook}. How it works for strings and branes,
is discussed in \ci{PavsicArena}--\ci{PavsicBook}. The existing
literature\,\ci{PavsicBook,Stueckelberg,Stueckelberg1}
shows that the Stueckelberg theory is a consistent
physical theory, and that the quantization of the Stueckelberg field is
not problematic at all.

Since the Stueckelberg theory admits tachyons, a question arises as to
whether it is not in conflict with causality. Because tachyons in some
reference frames are observed to move forward, and in some other frames
to move backwards in the time coordinate $x^0$, the arrangements with
tachyon emitters and absorbers are possible, such that they lead to
causal loops in $M_{1,3}$. Since the latter space is a subspace of $C$,
in most cases those loops will be merely projections of the lines
in $C$ onto $M_{1,3}$, and hence not true loops in $M_{1,3}$.
The $\tau$ and other coordinates of $C$ could be different, even if
$x^\mu$ coincided. Hence, there would be no paradox, because no ``change
of history" would be possible in such a case. Thus, the problem would
only be with the true loops in $C$, if they can occur. Then, one should
take into account that what propagates, are not classical, sharply
localized particles, but wave packets that always have a certain
width. When such a wave packet of a causal loop arrives at the critical point
in the past, a superposition of several possible histories would take place.
There would be no causal paradox, but only a `paradox' of several
co-existing histories\,\ci{PavsicBook,PavsicEverett,Deutsch},
which---according to the Everett many worlds
interpretation\,\ci{Everett,Everett1} of quantum mechanics---is not a paradox at all. In fact,
the existence of causal loops, if experimentally established, would
confirm the validity of the Everett interpretation of quantum mechanics,
and disprove the other interpretations.

The same form of an ultrahyperbolic equation in 16$D$ Clifford space, $C$,
holds for bradyons and tachyons. In order to experience a consistent
evolution of localized fields, an observer must live on a suitable light-like
hypersurface. All events on such hypersurface have the same value of
the evolution parameter $\tau$. As the $\tau$ changes, so does the hypersurface.
In other words, an observer is associated with a 1-parameter family of
light-like hypersurfaces in $C$, distinguished by values of $\tau$.
The families of hypersurfaces in $C$ that are not light-like cannot be
associated with observers, because the Cauchy problem for such hypersurfaces
is not well posed. An alternative possibility is that some observers are
associated with space-like or time-like 4-surfaces on which the initial data for the
Klein-Gordon field $\phi$ can be arbitrarily specified. Now the initial data
for the bradyonic field, given on a space-like 4-surface, include not only the
derivative of $\phi$ with respect to
one, but with respect to four time-like coordinates. For the tachyonic field,
the initial data are given on a time-like 4-surface, and include the derivatives
of the field with respect to four space-like coordinates.
Knowing such initial data, an observer can determine, although not uniquely,
the behavior of the field outside the 4-surface. In our opinion, such
indeterminism at the level of first quantized fields, should not be
considered as problematic, because in a more complete, second
quantized theory, fields are not well determined anyway.

The observed four-dimensionality of our spacetime implies that in
any higher-dimensional theory the extra dimensions must be either compactified
or the observed matter must be localized on a four-dimensional surface in
the higher-dimensional space. In the known examples where such
compactification/localization occurs it requires highly non-trivial dynamical
mechanisms that have been extensively studied in the literature.
Any dynamical resolution of
this issue must clearly lead to an effectively four-dimensional theory.
That, according to the above, cannot be invariant under the superluminal boosts.
In other words, any compactification/localizaton  mechanism will break
the invariance under the superluminal Lorentz transformations,
rendering such transformations irrelevant in four dimensions. Such a reasoning
would hold, if there would be only one four-dimensional surface, $M_{1,3}$, with
observers observing the localized matter on $M_{1,3}$. However,
the effective four-dimensional theory can be either in the spacetime $M_{1,3}$
of an observer ${\cal O}$, or in another spacetime ${\tl M}_{1,3}$ of an
observer ${\tl {\cal O}}$. The two spacetimes, $M_{1,3}$ and ${\tl M}_{1,3}$,
are subspaces of the higher-dimensional spacetime that we
started from, and can be transformed into each other by a superluminal boost
in that higher-dimensional space. If for the higher-dimensional space we
take $M_{4,4}$ of Sec.\,4, with coordinates $(x^\mu,{\tl x}^\mu)$,
$\mu=0,1,2,3$, then the subspace $M_{1,3}$ has coordinates
$x^\mu=(x^0,x^1,x^2,x^3)$, whereas the
subspace ${\tl M}_{1,3}$ has coordinates $x^0,x^1,{\tl x}^2,{\tl x}^3$.
Relative to
${\cal O}$, the observer ${\tl {\cal O}}$ is tachyonic, and vice versa.
Hence, bradyonic and tachyonic observers do not live in the same effective
four-dimensional spacetime. The intersection of their respective spacetime
is the two-dimensional spacetime, $M_{1,1}$, with coordinates $x^0$, $x^1$.

There is also a possibility that does not require a compactification
of extra dimensions, or a localization of the observed matter.
In the setup considered in Sec.\,4, we have a very special higher-dimensional
space, namely the Clifford space, $C$.
All the dimensions of $C$ are assumed to be observable even
at the macroscopic scale, because those dimensions are associated with
the position, size and orientation of the object. Therefore, the extra
dimensions of $C$ can be large, and need not be compactified at sufficiently
small scales, and be thus unobservable at macroscopic scales. If $C$ is curved,
then we have the gravity in $C$. In order to reproduce, \`a la Kaluza-Klein,
the usual, 4$D$ gravity in $M_{1,3}$ and the Yang-Mills interactions, 
there must be suitable isometries in $C$, given in terms of a set of
Killing vector fields. We expect that such isometries arise dynamically
in the presence of suitable extended sources, e.g., the branes in
$C$ \,\ci{PavsicMaxwellBrane}. How precisely this works, remains
to be investigated. The presence of isometries does not render
the extra dimensions of $C$, i.e., the object's size and orientation,
unobservable,  it only makes the theory to be
in agreement with the fact that effectively we have 4D gravity and
Yang-Mills interactions.

\section{Conclusion}

We have shown that tachyons are not in conflict with
special relativity and field theory, if those theories are properly extended.
It is well-known that special relativity can be extended to
encompass not only the subluminal, but also the superluminal reference frames,
and the transformations relating those frames. Then it turns out that
spacetime must be a complex 4-dimensional space, or a real 8-dimensional
space with neutral signature $(4,4)$. The Klein-Gordon or the Dirac
equation in such
space is ultrahyperbolic, and thus problematic, regardless of whether the field
is bradyonic or tachyonic. We have shown that the problem can be resolved
if we take the Cauchy data not on a 7-dimensional hypersurface, but on
a 4-dimensional surface that is space like for bradyonic and time like
for tachyonic fields. The Klein-Gordon or the Dirac equation also is
not problematic in the 16-dimensional Clifford space, $C$, where it
can be written in the form of the corresponding generalized
Stueckelberg like equation which describes localized propagating
bradyonic and tachyonic fields. 

Because bradyons in vacuum do not emit
\v Cerenkov radiation, also tachyons of the extended relativity
do not emit \v Cerenkov radiation in vacuum\,\ci{Recami,Recami1}.
This is a consequence of the postulated
symmetry between bradyons and tachyons. The expectation that such radiation
should accompany superluminal particles, is based on different theoretical
assumptions. Thus, according to the theory considered in this paper,
no \v Cerenkov radiation in the form of electron-positron pairs
is emitted by superluminal neutrinos.

\vs{6mm}

\centerline{\bf Acknowledgment}

\vs{2mm}

This work was supported by the Slovenian Research Agency.

\end{document}